\documentclass[a4paper,12pt]{article}
\pdfoutput=1

\usepackage{jheppub}

\usepackage{amsthm}

\usepackage[numbers,sort&compress]{natbib}
\usepackage{dsfont}
\usepackage{slashed}
\usepackage{color}
\usepackage{empheq}
\usepackage{xcolor}
\usepackage{comment}
\usepackage{bbold}

\usepackage{adjustbox}

\usepackage{soul}

\usepackage[english]{babel}

\usepackage[utf8x]{inputenc}

\usepackage{graphicx,epsfig}

\usepackage{tikz}

\usepackage{bashful}

\usepackage{subfigure}

\usepackage{setspace}

\usepackage{booktabs}

\usepackage{float}

\usepackage{nextpage}

\usepackage{pdfpages}

\usepackage{hypcap}

\usepackage[small,bf]{caption}

\usepackage{longtable}

\usepackage[makeroom]{cancel}
\usepackage{multirow}
\usepackage{microtype}

\usepackage{feynmp-auto-mod}

\usepackage{xparse}
\usepackage{etoolbox}

\usepackage{pgfplots}

\usepackage[subfigure]{tocloft}

\newcommand{\p}{\partial}

\renewcommand{\Re}{\mathrm{Re}}
\renewcommand{\Im}{\mathrm{Im}}

\renewcommand{\O}{\mathcal{O}}

\newcommand{\tr}{\mathrm{Tr}}

\newcommand{\nn}{\nonumber\\}

\newcommand{\q}{\mathsf{q}}

\newcommand{\msbar}{$\overline{\text{MS}}$}
\newcommand{\MeV}{\;\mathrm{MeV}}
\newcommand{\GeV}{\;\mathrm{GeV}}
\newcommand{\TeV}{\;\mathrm{TeV}}

\newcommand{\hc}{\mathrm{h.c.}}

\NewDocumentCommand{\Op}{ m m O{} o }{
	\O^{\ifblank{#3}{}{#3,}#2 }_{\IfNoValueTF{#4}{#1}{\substack{#1\\#4}}}
}
\NewDocumentCommand{\lwc}{ m m O{} o }{
	L^{\ifblank{#3}{}{#3,}#2 }_{\IfNoValueTF{#4}{#1}{\substack{#1\\#4}}}
}
\NewDocumentCommand{\dlwc}{ m m O{} o }{
	{\dot L}^{\ifblank{#3}{}{#3,}#2 }_{\IfNoValueTF{#4}{#1}{\substack{#1\\#4}}}
}
\NewDocumentCommand{\tlwc}{ m m O{} o }{\widetilde
	L^{\ifblank{#3}{}{#3,}#2 }_{\IfNoValueTF{#4}{#1}{\substack{#1\\#4}}}
}
\NewDocumentCommand{\cwc}{ m m O{} o }{
	C^{\ifblank{#3}{}{#3}#2 }_{\IfNoValueTF{#4}{#1}{\substack{#1\\#4}}}
}

\NewDocumentCommand{\tcwc}{ m m O{} o }{
	\widetilde C^{\ifblank{#3}{}{#3}#2 }_{\IfNoValueTF{#4}{#1}{\substack{#1\\#4}}}
}

\definecolor{darkgreen}{rgb}{0,0.5,0}
\definecolor{darkblue}{rgb}{0,0,0.5}
\definecolor{darkred}{rgb}{0.5,0,0}
\definecolor{beige}{rgb}{0.7,0.4,0.3}

\makeatletter
  \def\my@tag@font{\normalsize}
  \def\maketag@@@#1{\hbox{\m@th\normalfont\my@tag@font#1}}
  \let\amsmath@eqref\eqref
  \renewcommand\eqref[1]{{\let\my@tag@font\relax\amsmath@eqref{#1}}}
\makeatother

\newcommand{\abs}[1]{\left |  #1 \right |}

\newenvironment{myfmf}[1]
{\begin{fmffile}{#1}
\fmfcmd{%
  style_def wboson expr p =
  cdraw (wiggly p);
  shrink (1);
  cfill (arrow p);
  endshrink;
  enddef;}
\fmfcmd{%
  style_def momins expr p =
  drawarrow p;
  enddef;}
\fmfset{arrow_len}{2.5mm}
\fmfpen{0.75pt}
\fmfset{dot_size}{2pt}
  }
{
\end{fmffile}
}

\makeatletter
\renewcommand\paragraph{\@startsection{paragraph}{4}{\z@}%
  {-3.25ex\@plus -1ex \@minus -.2ex}%
  {1.5ex \@plus .2ex}%
  {\normalfont\normalsize\bfseries}}
\makeatother

\cftsetindents{section}{0em}{2em}
\cftsetindents{subsection}{2em}{3em}
\cftsetindents{subsubsection}{5em}{4em}

\allowdisplaybreaks[1]

\title{\boldmath Effective field theory interpretation of lepton magnetic and electric dipole moments}

\author[1]{Jason~Aebischer,}
\author[1]{Wouter~Dekens,}
\author[1]{Elizabeth~E.~Jenkins,}
\author[1]{Aneesh~V.~Manohar,}
\author[1]{Dipan~Sengupta,}
\author[1,2]{Peter~Stoffer}

\emailAdd{jaebischer@physics.ucsd.edu}
\emailAdd{wdekens@physics.ucsd.edu}
\emailAdd{ejenkins@ucsd.edu}
\emailAdd{amanohar@ucsd.edu}
\emailAdd{disengupta@physics.ucsd.edu}
\emailAdd{peter.stoffer@univie.ac.at}

\affiliation[1]{Department of Physics, University of California at San Diego, 9500 Gilman Drive,\\ La Jolla, CA 92093-0319, USA}
\affiliation[2]{University of Vienna, Faculty of Physics, Boltzmanngasse 5, 1090 Vienna, Austria}

\abstract{
We perform a model-independent analysis of the magnetic and electric dipole moments of the muon and electron. We give expressions for the dipole moments in terms of operator coefficients of the low-energy effective field theory (LEFT) and the Standard Model effective field theory (SMEFT). We use one-loop renormalization group improved perturbation theory, including the one-loop matching from SMEFT onto LEFT, and one-loop lepton matrix elements of the effective-theory operators. Semileptonic four-fermion operators involving light quarks give sizable non-perturbative contributions to the dipole moments, which are included in our analysis.
We find that only a very limited set of the SMEFT operators is able to generate the current deviation of the magnetic moment of the muon from its Standard Model expectation.
}

\numberwithin{equation}{section}

\begin{document}

	\mbox{} \\[-0.58cm] \mbox{} \hfill{} UWThPh 2021-2 \\[-1.12cm]

	\maketitle


	\begin{myfmf}{diags/diags}


\section{Introduction}
\label{sec:Introduction}

The electric and magnetic dipole moments of the electron and the muon are precision low-energy probes of the Standard Model (SM) of particle physics, and provide strong indirect constraints on physics beyond the Standard Model (BSM). The electric dipole moment is $CP$-odd, whereas the magnetic moment is $CP$-even. With the current experimental precision, the electron magnetic moment is sensitive not only to QED effects but also to hadronic contributions. Due to the larger muon mass, the muon magnetic moment is more sensitive to heavier mass scales, including hadronic and electroweak contributions, as well as BSM physics (see Ref.~\cite{Aoyama:2020ynm} for a comprehensive recent review). The current experimental value for the anomalous magnetic moment of the muon $a_\mu$~\cite{Bennett:2006fi,Abi:2021gix,Albahri:2021ixb,Albahri:2021kmg,Albahri:2021mtf} ($a_\ell = (g_\ell-2)/2,\ \ell=e,\mu$)
\begin{align}
	a_\mu^\mathrm{exp} &= 116\,592\,061(41) \times 10^{-11},
\end{align}
and the SM prediction~\cite{Aoyama:2020ynm,Aoyama:2012wk,Aoyama:2019ryr,Czarnecki:2002nt,Gnendiger:2013pva,Davier:2017zfy,Keshavarzi:2018mgv,Colangelo:2018mtw,Hoferichter:2019mqg,Davier:2019can,Keshavarzi:2019abf,Kurz:2014wya,Melnikov:2003xd,Colangelo:2015ama,Masjuan:2017tvw,Colangelo:2017qdm,Colangelo:2017fiz,Hoferichter:2018kwz,Gerardin:2019vio,Bijnens:2019ghy,Colangelo:2019uex,Blum:2019ugy,Colangelo:2014qya}
\begin{align}
	a_\mu^\mathrm{SM} &= 116\,591\,810(43) \times 10^{-11}\,,
\end{align}
are in tension at the level of $4.2\,\sigma$. Whether the difference
\begin{align}
	\label{eq:MuonDiscrepancy}
	\Delta a_\mu &= a_\mu^\mathrm{exp} - a_\mu^\mathrm{SM} = 251(59) \times 10^{-11}\,,
\end{align}
is a hint for new physics is not yet clear, but the E989 experiment at Fermilab is aiming to further reduce the experimental uncertainty to $0.14~\mathrm{ppm}$~\cite{Grange:2015fou}.
On the theory side, the latest determination of hadronic vacuum polarization (HVP) from lattice QCD~\cite{Borsanyi:2020mff} improves the uncertainties compared to previous lattice-QCD calculations, but it is in a $2.1\,\sigma$ tension with the value derived from the $e^+e^-\to\text{hadrons}$ cross section~\cite{Davier:2017zfy,Keshavarzi:2018mgv,Colangelo:2018mtw,Hoferichter:2019mqg,Davier:2019can,Keshavarzi:2019abf,Crivellin:2020zul,Keshavarzi:2020bfy,Malaescu:2020zuc,Colangelo:2020lcg}. Meanwhile, for the hadronic light-by-light contribution, lattice-QCD determinations~\cite{Blum:2019ugy,Chao:2021tvp} are in agreement with the phenomenological estimate~\cite{Aoyama:2020ynm}.

In the case of the electron anomalous magnetic moment, the agreement between experiment~\cite{Hanneke:2008tm},
\begin{align}
	a_e^\mathrm{exp} &= 1\,159\,652\,180.73(28) \times 10^{-12} \, ,
\end{align}
and the SM prediction crucially depends on the input value for the fine-structure constant $\alpha$. Here the two latest determinations based on Cesium and Rubidium atomic recoils
\begin{align}
\label{eq:alphaQED}
	\alpha^{-1}_\mathrm{QED,Cs} &= 137.035999046(27) \ \text{\cite{Parker:2018vye}} \, , \nn
	\alpha^{-1}_\mathrm{QED,Rb} &= 137.035999206(11) \  \text{\cite{Morel:2020dww}} \, ,
\end{align}
differ by more than $5\,\sigma$ and lead to SM predictions
\begin{align}
	\label{eq:1.6}
	a_e^\mathrm{SM,Cs} &= 1\,159\,652\,181.61(23) \times 10^{-12} \, , \nn
	a_e^\mathrm{SM,Rb} &= 1\,159\,652\,180.252(95) \times 10^{-12} \, ,
\end{align}
and the differences
\begin{align}
	\label{eq:ElectronDiscrepancy}
	\Delta a_e^\mathrm{Cs} &= a_e^\mathrm{exp} - a_e^\mathrm{SM,Cs} = - 0.88(36) \times 10^{-12} \, , \nn
	\Delta a_e^\mathrm{Rb} &= a_e^\mathrm{exp} - a_e^\mathrm{SM,Rb} = 0.48(30) \times 10^{-12} \, ,
\end{align}
which correspond to $-2.4\,\sigma$ or $+1.6\,\sigma$ discrepancies, respectively. The change in $a_\ell$ due to a change in $\alpha_\mathrm{QED}$ is dominated by the one-loop Schwinger correction $\alpha/(2\pi)$~\cite{Schwinger:1948iu}, which is the same for $a_e$ and $a_\mu$. However, the experimental uncertainty for $a_\mu$ is much larger than for $a_e$, and the theoretical prediction for the anomalous magnetic moment of the muon is not affected by the small difference in $\alpha_\mathrm{QED}$.
In the examples we consider in this paper, $\Delta a_\ell \propto m_\ell$, so it is useful to write the anomalous magnetic moment of the electron discrepancy as
\begin{align}
	\label{eq:Discrepancies2}
	\Delta a_e &= \frac{m_e}{m_\mu} \left[ {- 18 (7)^\mathrm{Cs} \atop \phantom{+}10(6)^\mathrm{Rb} } \times 10^{-11} \right] \,,
\end{align}
in order to compare BSM contributions to  $\Delta a_\mu$ and $\Delta a_e$ simultaneously. Dipole moments require a chirality change of the lepton, which can introduce an additional factor of the lepton mass, resulting in the ``naive scaling'' $\Delta a_\ell \propto m_\ell^2$~\cite{Giudice:2012ms}. A chiral enhancement results if this second mass factor is replaced by some heavy internal mass scale.

The electric dipole moments (EDMs) of leptons are $CP$-odd observables that are highly suppressed in the SM. Any measurement of an EDM would be a clear indication of BSM physics, and the current experimental bounds~\cite{Andreev:2018ayy,Bennett:2008dy}
\begin{align}\label{eq:electric}
	| d_e | &< 1.1 \times 10^{-29} \; e \text{--}\mathrm{cm} \qquad @\,  90\%\, \mathrm{CL} \, , \nn
	| d_\mu | &< 1.5 \times 10^{-19} \; e \text{--} \mathrm{cm} \qquad @\, 90\%\, \mathrm{CL} \,,
\end{align}
place strong constraints on new physics which is $CP$-violating.\footnote{The limit on $d_e$ also places an indirect constraint on $d_\mu$ that is slightly better than the direct experimental bound~\cite{Crivellin:2018qmi}.} The bound on the muon EDM could be improved by about three orders of magnitude in a proposed experiment at PSI~\cite{Adelmann:2021udj}.

In this paper, we investigate the possibility that the discrepancies, Eqs.~\eqref{eq:MuonDiscrepancy} and~\eqref{eq:ElectronDiscrepancy}, are due to BSM physics. While many models explaining the muon (and in some cases also the electron) $g-2$ discrepancy have been constructed, we instead analyze possible BSM contributions in a model-independent way using effective-field-theory (EFT) techniques.
EFT methods have been applied at the one-loop and even two-loop level to extract constraints from the EDM of the electron~\cite{Pruna:2017tif,Panico:2018hal}, but in the case of magnetic moments, the usual treatment only involves a discussion of the dipole operator itself. Recent studies investigated the connection to muon-collider~\cite{Buttazzo:2020eyl,Capdevilla:2021rwo} and muon-proton collider observables~\cite{Cheung:2021iev}. Logarithmic one-loop terms were included in Ref.~\cite{Buttazzo:2020eyl}.

In this work, we extend the available EFT analyses by exploring the lepton dipole moments at one-loop order including dimension-six operators. We consider two scenarios: (a)  BSM physics appears above the electroweak scale and can be analyzed using dimension-six operators in the SM effective field theory (SMEFT), and  (b)  BSM physics occurs between the electroweak and QCD scales, and can be analyzed using a low-energy effective field theory (LEFT).  Both effective field theories contain dipole operators that contribute to lepton dipole moments at tree level, so one trivial possibility is that the $g-2$ discrepancies are explained by a BSM contribution to the EFT dipole operator. In addition to this possibility, we take into account other operators that can contribute to the dipole moments at loop level.
Recent calculations have determined the one-loop anomalous dimensions in SMEFT~\cite{Jenkins:2013zja,Jenkins:2013wua,Alonso:2013hga,Alonso:2014zka} and LEFT~\cite{Jenkins:2017jig,Jenkins:2017dyc}, as well as the one-loop matching between the two theories~\cite{Dekens:2019ept}. We determine all operators which can contribute to $g-2$ including one-loop matching and perturbative integration of the renormalization group (RG) evolution up to second order in the anomalous dimensions.  We present our results in the spirit of Ref.~\cite{Aebischer:2018quc,Aebischer:2018csl,Aebischer:2020dsw} as master formulae in terms of LEFT, as well as SMEFT, Wilson coefficients.
We find that, although there are several operators that could provide an explanation if new physics arises below the electroweak scale, an explanation for the discrepancy based on other EFT operators arising from BSM physics heavier than the electroweak scale is highly constrained.
In this case, new physics would have to directly induce the dipole operator itself or certain semileptonic tensor operators, which can give significant contributions to $\Delta a_\ell$ through charm- or top-quark loops. Thus, while there are a plethora of new-physics models that aim to address $\Delta a_\ell$, within the EFT language, all of the models can be described by only a few operators.
We  end by showing how our results can be applied to specific BSM scenarios, using  the case of scalar leptoquarks as a representative example.

As the magnetic and electric dipole moments are related to the real and imaginary parts of the Wilson coefficients in SMEFT and LEFT, we also discuss the suppression of the imaginary parts relative to the real parts that is required to produce the $g-2$ anomalies while still being consistent with the electric dipole moment constraints in Eq.~\eqref{eq:electric}.


\section{Effective field theory and dipole moments}
\label{sec:EFT}

The low-energy effects of BSM physics above the electroweak scale can be described by an EFT that only contains the SM particles and respects the SM gauge group $SU(3)_c \times SU(2)_L \times U(1)_Y$, known as SMEFT~\cite{Weinberg:1979sa,Buchmuller:1985jz,Grzadkowski:2010es}.\footnote{We consider only the case where  the Higgs particle is part of an electroweak doublet as in the SM.  The case in which the Higgs boson is an $SU(2)_L$ singlet, known as Higgs Effective Field Theory (HEFT), is another possibility.} We limit our analysis to dimension-six operators. For operators of even higher dimension to be relevant, the BSM model would have to be tuned so that dimension-six contributions to $g-2$ are smaller than power-suppressed higher-dimension contributions.
If the BSM theory has light particles below the electroweak scale, but heavier than the muon mass, we can use a low-energy EFT that only contains light SM particles,  and is invariant under $SU(3)_c \times U(1)_\mathrm{em}$, known as LEFT.

Our sign convention for the $SU(3)_c \times SU(2)_L \times U(1)_Y$ covariant derivative is
\begin{align}
	D_\mu = \p_\mu + i g_3 T^A G_\mu^A + i g_2 t^I W_\mu^I + i g_1 \mathsf{y} B_\mu \, ,
\end{align}
with $SU(3)_c$ generators $T^A$,  $SU(2)_L$ generators $t^I = \tau^I / 2$, and the $U(1)_Y$ hypercharge generator $\mathsf{y}$. Our sign convention for the $SU(3)_c \times U(1)_\mathrm{em}$ covariant derivative is
\begin{align}
	D_\mu = \p_\mu + i g T^A G^A_\mu + i e Q A_\mu \, ,
\end{align}
where the QED coupling constant is positive, $e = |e|$, and the eigenvalues of the electric charge operator $Q$ are $\q_e = -1$, $\q_u = 2/3$, $\q_d = -1/3$ for the charged leptons, up-type quarks, and down-type quarks, respectively.

The results for the SMEFT and LEFT running and matching calculations have been implemented in publicly available codes~\cite{Celis:2017hod,Aebischer:2017ugx,Aebischer:2018bkb,Fuentes-Martin:2020zaz},\footnote{An overview of these codes can be found in Ref.~\cite{Brivio:2019irc}. In the course of our analysis, we discovered bugs in the public codes {\tt wilson}~\cite{Aebischer:2018bkb} and {\tt DSixTools}~\cite{Celis:2017hod}, and we are in communication with the authors of these codes.
} which we partly use for our numerical analysis. As the weak interactions are a chiral gauge theory, the finite parts of one-loop diagrams depend on the specific scheme chosen for $\gamma_5$. For example, the Fierz identity
\begin{align}
\label{eq:Fierz}
\bar A \gamma^\mu P_L B \ \bar C \gamma_\mu P_R D &= - 2\, \bar A P_R D \ \bar C P_L B
\end{align}
holds in $d=4$ dimensions. Away from $d=4$ dimensions, the difference of the two sides of Eq.~\eqref{eq:Fierz} is an evanescent operator, which has matrix elements proportional to $\left( d-4 \right)$, and gives a non-zero finite contribution in one-loop matrix elements if the loop integral is divergent. We will use the SMEFT operator basis in Ref.~\cite{Grzadkowski:2010es}, the LEFT operator basis in
Ref.~\cite{Jenkins:2017jig}, and follow the conventions for evanescent operators and $\gamma_5$ explained in Ref.~\cite{Dekens:2019ept}.

The anomalous magnetic moment and electric dipole moment of a lepton are defined through the vertex function
\begin{align}
	\label{eq:VertexFunction}
	\begin{gathered}
		\begin{fmfgraph*}(60,50)
			\fmftop{t1} \fmfbottom{b1,b2}
			\fmf{quark,label=$\ell$,label.side=left,tension=2}{b1,v1}
			\fmf{quark,label=$\ell$,label.side=left,tension=2}{v1,b2}
			\fmf{photon,label=$\gamma$,tension=3}{t1,v1}
			\fmfblob{6mm}{v1}
		\end{fmfgraph*}
	\end{gathered}
	\hspace{1cm} = \hspace{0.5cm} (-ie \q_e)\, \bar u(p') \Gamma^\mu(p,p') u(p) \, ,
\end{align}
which is decomposed in terms of Lorentz-invariant form factors as
\begin{align}
	\Gamma^\mu(p,p') &= \gamma^\mu F_E(k^2) + i \frac{\sigma^{\mu\nu} k_\nu}{2m_\ell} F_M(k^2) + \frac{\sigma^{\mu\nu} k_\nu}{2m_\ell} \gamma_5 F_D(k^2) + \frac{k^2 \gamma^\mu - k^\mu \slashed{k} }{m_\ell^2} \gamma_5 F_A(k^2) \, ,
\end{align}
where $k=p'-p$ is the incoming photon momentum. The mass $m_\ell$ is the physical (pole) lepton mass and the external lepton states are on shell, $p^2 = {p'}^2 = m_\ell^2$.
The vertex function includes external-leg corrections on the photon and fermion lines. The prefactor $e$ in Eq.~\eqref{eq:VertexFunction} is fixed by the renormalization condition $F_E(0) = 1$, and corresponds to the static on-shell electric charge $e_\mathrm{QED}$, measured, for example, using Coulomb's law at large distances.  The two experimental values for $\alpha_\mathrm{QED} \equiv e_\mathrm{QED}^2/(4\pi) $ are given in Eq.~\eqref{eq:alphaQED}.

The anomalous magnetic moment is defined as the limit of the magnetic (or Pauli) form factor at zero momentum transfer:
\begin{align}
	a_\ell = \frac{1}{2}(g_\ell-2) = F_M(0) \, ,
\end{align}
and the electric dipole moment is given by the limit of the electric dipole form factor at zero momentum transfer:
\begin{align}
	d_\ell = -\frac{e_\mathrm{QED} \q_e}{2m_\ell} F_D(0) \, .
\end{align}
The anapole moment $F_A(0)$ is not an observable~\cite{Musolf:1990sa}; it can be absorbed into four-fermion operators using the equations of motion.

To illustrate the LEFT description of magnetic dipole moments, consider the SM one-loop electroweak contribution to the anomalous magnetic dipole moments~\cite{Fujikawa:1972fe}
\begin{align}
	\label{eq:aEW}
	a_\ell^{\text{EW}}
	&= \frac{G_F m_\ell^2}{12 \sqrt{2}\, \pi^2} \left[ 3-4 \sin^2\theta_W + 8 \sin^4 \theta_W \right] + \O(v^{-4}) \, .
\end{align}
The terms in the LEFT Lagrangian that we need for this paper, which will allow us to describe both SM and BSM contributions, are
\begin{align}
	\label{eq:SMFourFermiA}
	\mathcal{L} &= \Big[ \lwc{e\gamma}{}[][pr]  (\bar e_{Lp}   \sigma^{\mu \nu} e_{Rr})\, F_{\mu \nu} +  \lwc{ee}{RR}[S][prst] (\bar e_{Lp}   e_{Rr})(\bar e_{Ls} e_{Rt}) \nn
		&\quad +   \lwc{eu}{RL}[S][prst] (\bar e_{Lp} e_{Rr})(\bar u_{Rs} u_{Lt})  +   \lwc{ed}{RL}[S][prst] (\bar e_{Lp} e_{Rr})(\bar d_{Rs} d_{Lt}) \nn
		&\quad +   \lwc{eu}{RR}[S][prst] (\bar e_{Lp} e_{Rr})(\bar u_{Ls} u_{Rt})  +   \lwc{ed}{RR}[S][prst] (\bar e_{Lp} e_{Rr})(\bar d_{Ls} d_{Rt}) \nn
		&\quad +   \lwc{eu}{RR}[T][prst] (\bar e_{Lp} \sigma^{\mu \nu}  e_{Rr})(\bar u_{Ls} \sigma_{\mu \nu} u_{Rt})  +   \lwc{ed}{RR}[T][prst] (\bar e_{Lp} \sigma^{\mu \nu}  e_{Rr})(\bar d_{Ls} \sigma_{\mu \nu} d_{Rt}) + \hc \Big] \nn
		&\quad +  \lwc{ee}{LR}[V][prst] (\bar e_{Lp}  \gamma^\mu e_{Lr})(\bar e_{Rs} \gamma_\mu e_{Rt})  \, ,
\end{align}
with implicit sums over the flavor (generation) indices $p,r,s,t$. We will denote the generation indices by $1,2,3$, or equivalently by $e,\mu,\tau$ for leptons, and $u,c,t$ for up-type quarks. For down-type quarks, the generation index is the weak-eigenstate index $d^\prime, s^\prime, b^\prime$ and is related to the mass eigenstate index $d,s,b$ by the CKM matrix, as discussed in Ref.~\cite[\S 2.2]{Jenkins:2017jig}.
We use the convention that the coefficients satisfy the same symmetry relations as the operators, e.g.,
\begin{align}
\label{eq:symmetry}
\lwc{ee}{LR}[V][prst] &= \lwc{ee}{LR*}[V][rpts] \, , \qquad
\lwc{ee}{RR}[S][prst] = \lwc{ee}{RR}[S][stpr]  \, ,
\end{align}
and both contributions are included in the flavor index sum in Eq.~\eqref{eq:SMFourFermiA}.
$ \lwc{e\gamma}{}[][]$ is the coefficient of a dimension-five operator, while the others are coefficients of dimension-six operators. The corresponding operators in the LEFT Lagrangian Eq.~\eqref{eq:SMFourFermiA} are denoted by $\Op{e\gamma}{}[][pr] $, $ \Op{ee}{LR}[V][prst]$, etc.

In constructing the LEFT Lagrangian induced by the SM, there is a tree-level contribution to the four-fermion operator $ \Op{ee}{LR}[V][prst]$ due to $Z$ exchange,
\begin{align}
	\label{eq:SMFourFermi}
	\lwc{ee}{LR}[V][prst] 	& = \delta_{pr} \delta_{st} \frac{4 G_F}{\sqrt 2}  \left(1- 2 \sin^2 \theta_W\right) \sin^2 \theta_W + \O(v^{-4})  \, ,
\end{align}
and a finite matching contribution to the dipole operator $\Op{e\gamma}{}[][pr] $ at one-loop,
\begin{align}
\label{eq:SMDipole}
	\lwc{e\gamma}{}[][p r] &= \delta_{p r} \frac{e \q_e G_F m_l}{48 \sqrt 2 \, \pi^2} \left(-3  - 8 \sin^2 \theta_W + 16 \sin^4 \theta_W  \right) + \O(v^{-4}) \, .
\end{align}

The total one-loop electroweak contribution to the anomalous magnetic moment is given by the sum of the tree-level contribution from the dipole operator Eq.~\eqref{eq:SMDipole} and the one-loop contribution of the $V,LR$ operator Eq.~\eqref{eq:SMFourFermi} from the graph shown in Fig.~\ref{fig:1}; the two contributions sum to  Eq.~\eqref{eq:aEW}.
\begin{figure}
\begin{center}
\includegraphics[height=2.5cm]{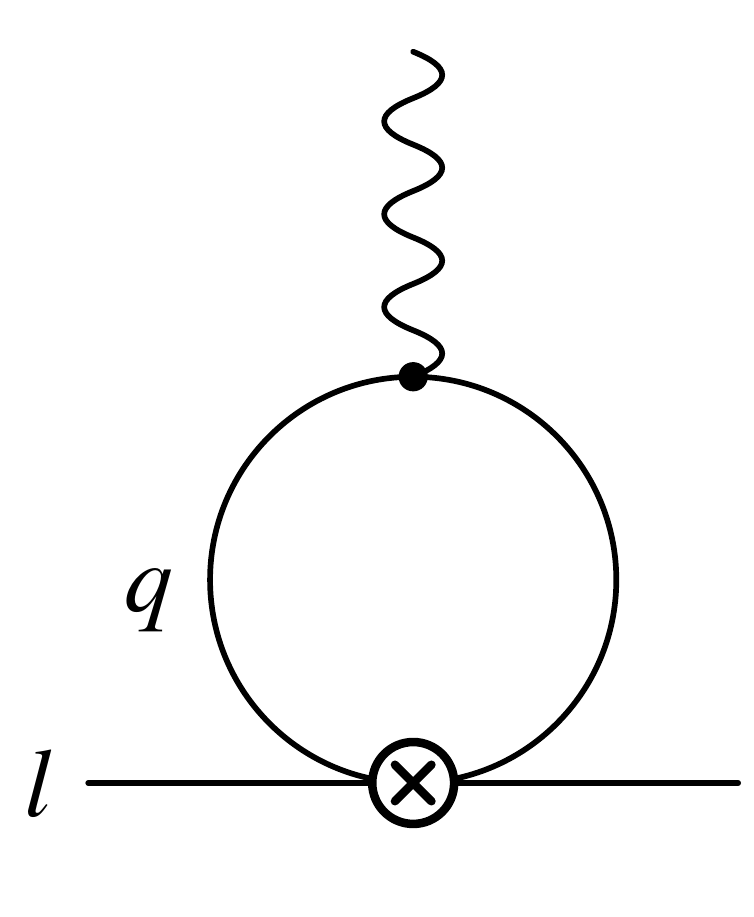}
\end{center}
\caption{\label{fig:1} One-loop contribution to the dipole moment from four-fermion operators.}
\end{figure}
While the $W$ contribution only enters as a direct matching to the dipole operator, numerically about $60\%$ of the $Z$ contribution comes from the LEFT loop and $40\%$ from the direct matching to the dipole. For BSM contributions, one generally expects an analogous situation: some effects only appear as a direct matching onto the dipole operators, while other contributions can indirectly feed into $g-2$ through EFT loops. We can evaluate the complete RG-improved BSM contribution since one-loop running, matching, and matrix elements are now available.

Lepton dipole  moments are predictions of the SM that only depend on the experimental inputs of particle masses and coupling constants. In the EFT, the dipole moments also depend on higher-dimension operator coefficients. The dipole moments get tree-level contributions from the dimension-five dipole operators as well as loop contributions from dimension-six interactions in the LEFT from the graph shown in Fig.~\ref{fig:1}. The dipole moments therefore depend on Wilson coefficients that are free parameters in the EFT, making it difficult to correlate $g-2$ with other observables in general. This feature explains why, for the magnetic moments, the EFT language is mostly employed only up to tree level, i.e., models are matched directly onto the dipole operators. We will show that the EFT approach beyond leading order is nevertheless useful to systematically organize viable mechanisms that can explain the $g-2$ discrepancies and to explicitly distinguish logarithmically enhanced contributions.

As is well known, the SM electroweak contribution to the muon $g-2$ up to two loops~\cite{Czarnecki:2002nt,Gnendiger:2013pva}
\begin{align}
	\label{eq:amuEWTwoLoops}
	a_\mu^\mathrm{EW} &= (153.6 \pm 1.0) \times 10^{-11}\,,
\end{align}
is comparable in magnitude to the discrepancy~Eq.~\eqref{eq:MuonDiscrepancy}, which already implies that heavy BSM physics needs an enhancement mechanism to be a viable explanation~\cite{Giudice:2012ms,Crivellin:2018qmi}. In the case of the electron $g-2$, both the differences Eq.~\eqref{eq:ElectronDiscrepancy} and the uncertainties are much larger than the SM electroweak contribution~\cite{Jegerlehner:2017gek},
\begin{align}
	a_e^\mathrm{EW} &= (30.53 \pm 0.23) \times 10^{-15} \,.
\end{align}


\section{LEFT interpretation}\label{sec:MatrixElement}

At tree level, the magnetic and electric dipole moments $a_\ell$ and $d_\ell$ are related to the real and imaginary parts of $\lwc{e\gamma}{}[][] $ and are conventionally given using different units, with $a_\ell$ dimensionless and $d_\ell$ in $e\text{--}\text{cm}$,
\begin{align}
- \frac{4 m_\ell }{e \mathsf{q}_e} \lwc{e\gamma}{}[][\ell \ell]
& = a_\ell +  \frac{2 m_\ell}{e_\mathrm{QED} \mathsf{q}_e}  i d_\ell \,, \nn
(6.750 \times 10^{-3} \GeV)\ \lwc{e\gamma}{}[][e e] \, & =   a_e - i  \left(5.2 \times 10^{10} \right)\frac{d_e}{1\,e\text{--}\text{cm}} \, , & \ell=e \, , \nn
(1.396 \GeV)\ \lwc{e\gamma}{}[][\mu \mu]\,  & =  a_\mu - i  \left(1.1 \times 10^{13} \right)\frac{d_\mu}{1\,e\text{--}\text{cm}}  \, , & \ell=\mu \, .
\label{3.1}
\end{align}
If $\Re( \lwc{e\gamma}{}[][ee] )$ explains the electron anomalous magnetic moment discrepancy $\Delta a_e^\mathrm{Cs,Rb}$,  $\Im( \lwc{e\gamma}{}[][ee]  )$ has to be suppressed relative to the real part by about $10^{-6}$ to satisfy the bound on the electric dipole moment $d_e$ given in Eq.~\eqref{eq:electric}. No suppression is needed for the muon.

At one-loop order, $a_\ell$ is
\begin{align}
	\label{eq:amuLEFT}
	a_\ell  &= \frac{\alpha \q_e^2}{2\pi} -4\frac{m_\ell}{e \q_e}\Re\, \lwc{e\gamma}{}[][\ell\ell] (\mu) \left\{ 1 - \frac{\alpha \q_e^2}{4\pi} \left[ 2 + 5 \log\left( \frac{\mu^2}{m_\ell^2} \right) \right] \right\} + a_{\ell}^{4\ell} + a^{2\ell 2q}_{\ell} + \O(\lwc{e\gamma}{2}[][])  \, ,
\end{align}
where $e$ now denotes the renormalized \msbar{} coupling, but $m_\ell$ is still the physical pole mass of the lepton.
The first term is the famous QED one-loop contribution computed by Schwinger~\cite{Schwinger:1948iu}. The QED contribution is known to five loops~\cite{Aoyama:2012wk}. The second term is the tree-level contribution of the dipole operator $\Op{e\gamma}{}[] $, together with its one-loop correction from the graphs in Fig.~\ref{fig:correction}.
\begin{figure}[t]
\begin{center}
\includegraphics[width=3cm]{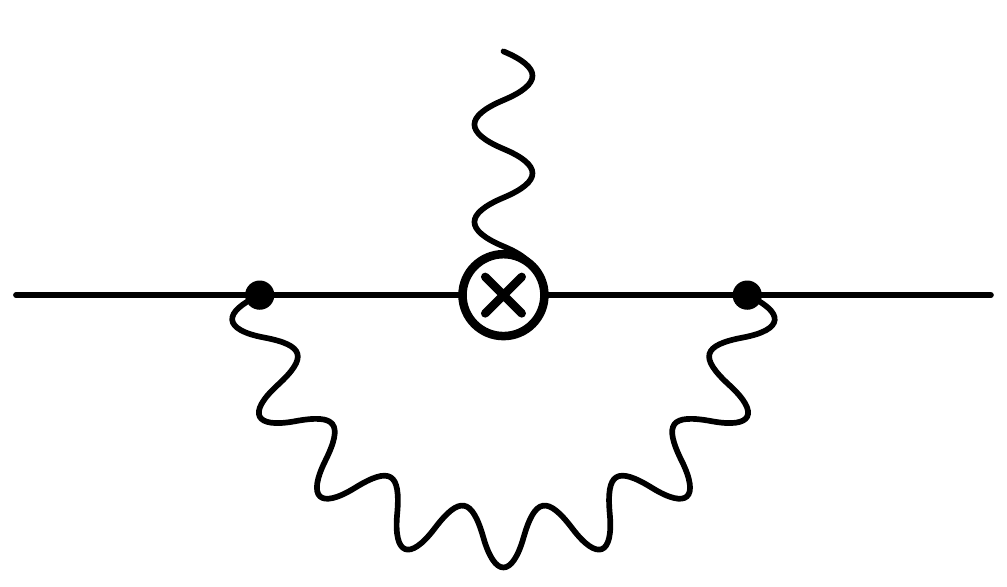} \hspace{1cm}
\includegraphics[width=3cm]{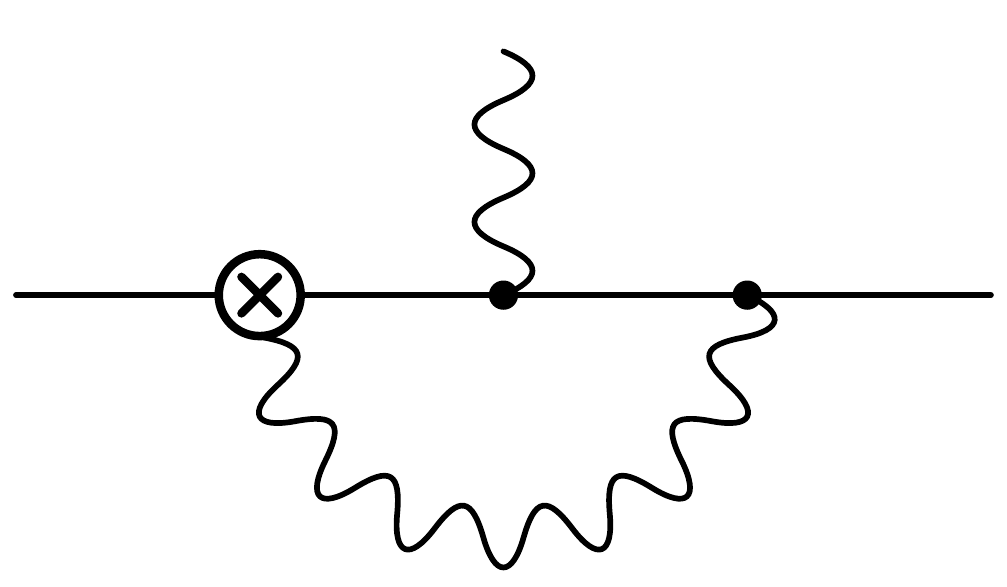} \hspace{1cm}
\includegraphics[width=3cm]{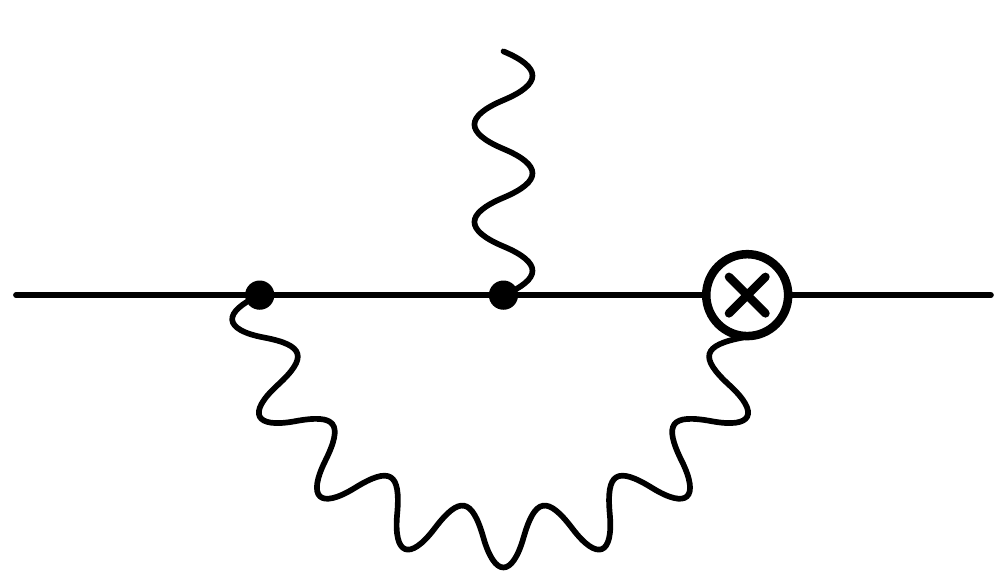}
\end{center}
\caption{One-loop correction to the matrix element of $\lwc{e\gamma}{}[][] $. There is also a wave-function correction, which is not shown.}
\label{fig:correction}
\end{figure}
The $a_{\ell}^{4\ell}$ contribution arises from one-loop insertions of four-lepton operators. In the \msbar{} scheme, and using the conventions of Ref.~\cite{Dekens:2019ept} for evanescent operators,
it is given by
\begin{align}\label{eq:amu4l}
 a_{\ell}^{4\ell }
 &= -m_\ell\sum_{p=e,\mu,\tau} \frac{ m_{l_p}}{4\pi^2} \ \Re\, \lwc{ee}{LR}[V][\ell p p \ell] (\mu)  + m_\ell\sum_{p=e,\mu,\tau} \frac{ m_{l_p}}{4\pi^2} \log \bigg( \frac{\mu^2}{m_{l_p}^2} \bigg) \Re\, \lwc{ee}{RR}[S][\ell p p \ell] (\mu)    \, ,
\end{align}
where $\mu$ denotes the renormalization scale, and we have used the symmetry of the Wilson coefficients Eq.~\eqref{eq:symmetry}.
The corresponding expressions for the electric dipole moment are given by the imaginary parts using the relative normalization in Eq.~\eqref{3.1} between $a_\ell$ and $d_\ell$. In the SM, the scalar Wilson coefficients $\lwc{ee}{RR}[S][]$ are proportional to Yukawa couplings and only appear as dimension-eight effects~\cite{Jenkins:2017jig}. The same remains true in the SMEFT at one loop~\cite{Dekens:2019ept}. However, these Wilson coefficients could be present in the LEFT if there is BSM physics below the weak scale. On the other hand, $\lwc{ee}{LR}[V][]$ is already present in the SM due to the one-loop electroweak contribution~\eqref{eq:SMFourFermi}.

Finally, $a^{2\ell 2q}_{\ell}$ in Eq.~\eqref{eq:amuLEFT} results from semileptonic tensor operators given in Eq.~\eqref{eq:SMFourFermi}. In perturbation theory and in the scheme of Ref.~\cite{Dekens:2019ept}, only logarithmic one-loop contributions are generated,
\begin{align}\label{eq:a2l2qPert}
 a_{\ell}^{2\ell 2q}
&= -m_\ell \sum_p \frac{N_c \mathsf{q}_u  m_{p}}{\mathsf{q}_e \pi^2 }   \log \bigg( \frac{\mu^2}{m_{p}^2} \biggr) \Re\, \lwc{eu}{RR}[T][\ell \ell p p] (\mu) \nn
	&\quad -m_\ell \sum_p \frac{N_c \mathsf{q}_d  m_{p}}{\mathsf{q}_e \pi^2 }   \log \bigg( \frac{\mu^2}{m_{p}^2} \biggr) \Re\, \lwc{ed}{RR}[T][\ell \ell p p] (\mu)   \, ,
\end{align}
where the sums on $p$ are over {\sl heavy} active quark flavors at the scale $\mu$. The light-quark matrix elements at low energies have to be evaluated non-perturbatively~\cite{Dekens:2018pbu}, and are not included in  Eq.~\eqref{eq:a2l2qPert}.  They give a dipole contribution
\begin{align}\label{eq:a2l2q}
	a_\ell^{2l2q,uds} &= 8c_T \frac{m_\ell}{ \q_e} \frac{F^2_\pi}{\Lambda_\chi}  \Re \left[ \q_u \lwc{eu}{RR}[T][\ell\ell uu] + \q_d \lwc{ed}{RR}[T][\ell\ell dd] + \q_d \lwc{ed}{RR}[T][\ell\ell ss]\right] \, ,
\end{align}
instead of the perturbative contribution Eq.~\eqref{eq:a2l2qPert}.
Here $\Lambda_\chi = 4\pi F_\pi$ is the scale of chiral symmetry breaking, with $F_\pi = 92\MeV$ the pion decay constant. $c_T$ is a non-perturbative parameter, of order unity by naive dimensional analysis~\cite{Manohar:1983md,Gavela:2016bzc}, and depends on the renormalization scale $\mu$ of the operators, which is chosen to be $2\GeV$.
In terms of the chiral low-energy constant $\Lambda_1$ of Ref.~\cite{Cata:2007ns},
\begin{align}
	c_T = \frac{4\pi}{F_\pi} \Lambda_1\, .
\end{align}
Lattice-QCD input from Ref.~\cite{Baum:2011rm} for the pion tensor charge together with vector-meson saturation~\cite{Ecker:1988te} lead to $c_T \approx -1.0(2)$, where the error does not include model uncertainties, see Refs.~\cite{Dekens:2018pbu,Hoferichter:2018zwu}.

\begin{figure}
\begin{center}
\includegraphics[width=3cm]{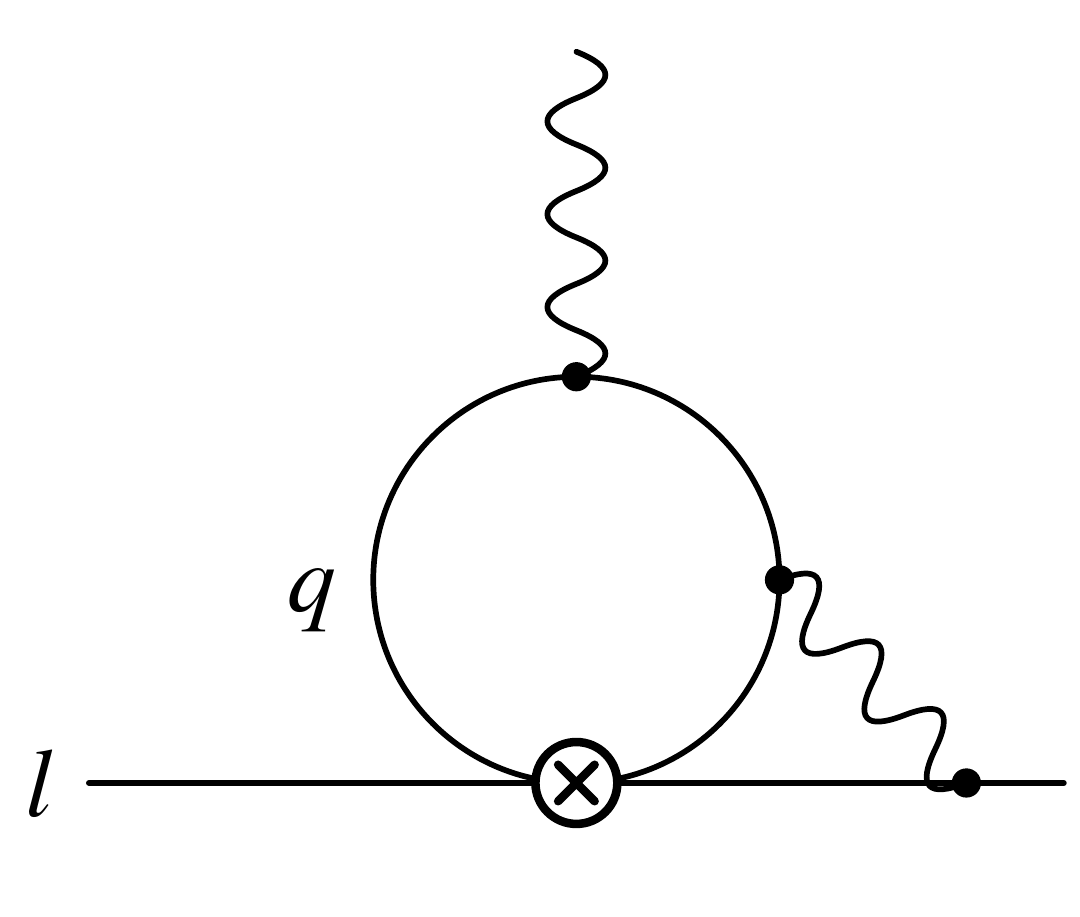} \hspace{1cm}
\includegraphics[width=3cm]{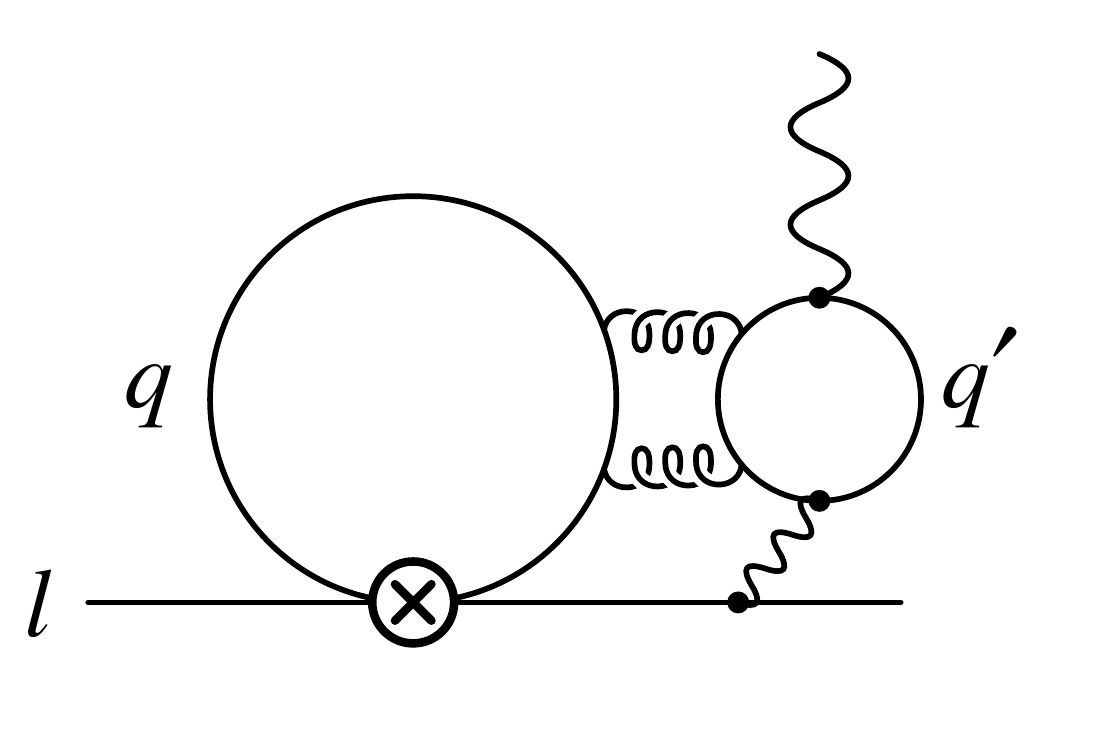}
\end{center}
\caption{Contributions of the semileptonic scalar operators to the lepton dipole. The right-hand diagram starts at four-loop order in perturbation theory. However, for light quarks, it gives a non-perturbative contribution to the dipole that is not suppressed by loop factors in comparison to the left-hand diagram.}
\label{fig:emmixing}
\end{figure}
The scalar operators $\lwc{ed}{RR}[S][ ]$, $\lwc{eu}{RR}[S][ ]$ mix into the corresponding tensor operators $\lwc{ed}{RR}[T][ ]$, $\lwc{eu}{RR}[T][ ]$ due to electromagnetic corrections, which subsequently contribute to the dipole via penguin graphs, as shown in Fig.~\ref{fig:emmixing}. This scalar--tensor mixing means that there are also non-perturbative contributions to the dipole from $\lwc{eu}{RR}[S][\ell \ell uu ]$, $\lwc{ed}{RR}[S][\ell \ell dd ]$, $\lwc{ed}{RR}[S][\ell \ell ss ]$ similar to Eq.~\eqref{eq:a2l2q}, but with an additional $\alpha \mathsf{q}_q \mathsf{q}_e/(4\pi) $ electromagnetic suppression due to photon exchange between the quark and lepton lines, from the left-hand graph in Fig.~\ref{fig:emmixing}. We will use the estimate
\begin{align}\label{eq:a2l2qS}
	a_\ell^{2l2q,uds,S} &= 8c_S \frac{\alpha}{4\pi} {m_\ell} \frac{F^2_\pi}{\Lambda_\chi}  \Re \left[ \q_u^2 \lwc{eu}{RR}[S][\ell\ell uu] + \q_d^2 \lwc{ed}{RR}[S][\ell\ell dd] + \q_d^2\lwc{ed}{RR}[S][\ell\ell ss]\right] 	\,,
\end{align}
where $c_S$ is a non-perturbative constant of order unity, and depends on $\mu$, which is chosen to be $2\GeV$.\footnote{In chiral perturbation theory, the terms in Eq.~\eqref{eq:a2l2qS} arise from structures $\propto\tr(Q_L^2L U^\dagger )  \O_{e\gamma}$ or $\tr(Q_LLQ_R U^\dagger )  \O_{e\gamma}$, where $U$ is the matrix containing the pions, $Q_{L,R} = {\rm diag}( \q_u, \q_d, \q_d)$, and $L = {\rm diag}( \lwc{eu}{RR}[S][\ell\ell uu] , \lwc{ed}{RR}[S][\ell\ell dd] , \lwc{ed}{RR}[S][\ell\ell ss] )$. There also exist structures $\propto\tr(L U^\dagger )$Tr$  (Q_L^2 ) \O_{e\gamma}$ from diagrams like the one in the right-hand panel of Fig.~\ref{fig:emmixing}, which give rise to a different $\q_{u,d}$ dependence. We neglect them for simplicity, as these contributions, like those in Eq.~\eqref{eq:a2l2qS}, have an electromagnetic suppression and are very small. }

At a scale of $\mu = 2\GeV$, the charm quark is an active degree of freedom. While non-perturbative effects in its contribution may not be negligible, we use perturbation theory, Eq.~\eqref{eq:a2l2qPert}, for a rough estimate of the charm contribution:
\begin{align}
	\label{eq:CharmLoop}
	a_\ell^{2l2q,c,T} \sim - c_T^{(c)} \frac{N_c m_\ell m_c \q_u}{\pi^2 \q_e} \log\left( \frac{\mu^2}{m_c^2} \right) \Re \lwc{eu}{RR}[T][\ell\ell cc](\mu) \, ,
\end{align}
where $c_T^{(c)}$ is expected to be of order unity, and is a reminder that there are non-perturbative contributions to the charm-quark loop. For the scalar charm-quark semileptonic operator, we will use the estimate
\begin{align}
	\label{eq:CharmLoopS}
	a_\ell^{2l2q,c,S} \sim - c_S^{(c)}\frac{\alpha}{4\pi} \frac{N_c m_\ell m_c \q_u^2}{\pi^2} \log\left( \frac{\mu^2}{m_c^2} \right) \Re \lwc{eu}{RR}[S][\ell\ell cc](\mu) \, ,
\end{align}
where the additional $\alpha \mathsf{q}_u \mathsf{q}_e/(4\pi)$ suppression relative to Eq.~\eqref{eq:CharmLoop} again comes from the photon exchange in the left-hand diagram in Fig.~\ref{fig:emmixing}.

The $(\overline L R)(\overline R L)$ operators $\lwc{ed}{RL}[S][ ]$, $\lwc{eu}{RL}[S][ ]$ also can have non-perturbative contributions from graphs where there is photon exchange between the quark and lepton lines. We will use the estimates Eq.~\eqref{eq:a2l2qS} and Eq.~\eqref{eq:CharmLoopS} with the non-perturbative constants replaced by $\widetilde c_S$ and $\widetilde c_S^{(c)}$, respectively. The non-perturbative contribution from semileptonic vector operators requires, in addition to the photon exchange, a helicity flip on the lepton line, with a resultant $m_\ell$ chiral suppression. As a result, their contribution to $a_\ell$ is suppressed relative to the scalar operators, and can be neglected.

In Eq.~\eqref{eq:amuLEFT}, we have neglected terms quadratic in the dipole-operator coefficients~\cite{Jenkins:2017dyc,Jenkins:2017jig}, which in principle appear at dimension six as well. Second-order contributions from two flavor-changing dipoles are negligible due to constraints on flavor-violating decays of charged leptons. Since the flavor-diagonal dipole operators contribute at tree level to the dipole moments, second-order dipole contributions are much smaller than the first-order tree-level contributions, and can be neglected.

In order to interpret the $g-2$ discrepancies in terms of new physics, we subtract from Eq.~\eqref{eq:amuLEFT} the corresponding expression in the SM: the Schwinger term drops out, and as we only keep terms linear in the Wilson coefficients, all LEFT coefficients can be replaced by their purely BSM contribution.

Let us evaluate Eq.~\eqref{eq:amuLEFT} numerically in order to estimate the size of the relevant Wilson coefficients needed to explain the anomalies. Using coefficients renormalized at $\mu=2\GeV$, one finds:
\begin{align}\label{eq:amuLEFTnum}
&\Delta a_{\ell}^{2\GeV} =\frac{m_\ell}{m_\mu}\Re\Bigg[ {1.35_\mu \atop 1.31_e} \, \lwc{e\gamma}{}[][\ell\ell]\, \times 1\GeV^{-1} \nn
&\quad -  1.4 \times 10^{-6}  \,\lwc{ee}{LR}[V][\ell ee \ell] - 2.8\times 10^{-4}\,\lwc{ee}{LR}[V][\ell \mu \mu \ell]- 4.8\times 10^{-3}\,\lwc{ee}{LR}[V][\ell \tau \tau \ell]\nn
&\quad+2.3\times 10^{-5}\,\lwc{ee}{RR}[S][\ell ee \ell]+1.7\times 10^{-3}\,\lwc{ee}{RR}[S][\ell \mu \mu \ell]+1.1\times 10^{-3}\,\lwc{ee}{RR}[S][\ell \tau \tau \ell]\nn
&\quad+2.1\times 10^{-3} c_T \left(\lwc{ed}{RR}[T][\ell\ell dd ]+\lwc{ed}{RR}[T][\ell\ell ss ] -2\,\lwc{eu}{RR}[T][\ell\ell uu]\right) + 2.7 \times 10^{-2} c_T^{(c)} \lwc{eu}{RR}[T][\ell\ell cc] \nn
&\quad+4.1\times 10^{-7} c_S \left(\lwc{ed}{RR}[S][\ell\ell dd ]+\lwc{ed}{RR}[S][\ell\ell ss ] +4 \,\lwc{eu}{RR}[S][\ell\ell uu]\right) - 1.1 \times 10^{-5} c_S^{(c)} \lwc{eu}{RR}[S][\ell\ell cc] \nn
&\quad+4.1\times 10^{-7}\widetilde c_S \left(\lwc{ed}{RL}[S][\ell\ell dd ]+\lwc{ed}{RL}[S][\ell\ell ss ] +4 \,\lwc{eu}{RL}[S][\ell\ell uu]\right) - 1.1 \times 10^{-5} \widetilde c_S^{(c)} \lwc{eu}{RL}[S][\ell\ell cc]  \Bigg] \times 1\GeV^2 \, .
\end{align}
For definiteness, we take the \msbar{} gauge couplings at the scale $\mu=M_Z$ as input~\cite{Zyla:2020zbs} and evolve them to $2\GeV$ using the one-loop RG.
For two generations of up- and down-type quarks and three lepton generations, the LEFT Lagrangian has 1176 $CP$-even baryon- and lepton-number-preserving terms at dimension six, and 25 at dimension five. Only 10 and 1 of these, respectively, contribute to each $\Delta a_\ell$.
Without any assumption on the flavor structure of BSM physics, the muon and electron dipole moments are essentially uncorrelated: 9 of the 11 free parameters in each $\Delta a_e$ and $\Delta a_\mu$ are unique and only the two parameters $\lwc{ee}{RR}[S][e\mu \mu e]$ and $\lwc{ee}{LR}[V][e \mu \mu e]$ enter both electron and muon dipole moments due to the symmetry relation Eq.~\eqref{eq:symmetry}.

Assuming  $\lwc{e\gamma}{}[][]\sim\Lambda^{-1}$, $\lwc{ee,ed,eu}{}[][]\sim\Lambda^{-2}$ where $\Lambda$  is the scale of new physics,  the largest impact on $a_\ell$ is from the  dimension-five operator $\lwc{e\gamma}{}[][]$. However, in many models $\lwc{e\gamma}{}[][]$ is induced at dimension-six, $\lwc{e\gamma}{}[][] \sim v/\Lambda^2$, and its contribution scales like the four-fermion operators. The most important four-fermion operators are the leptonic current-current operator $\lwc{ee}{LR}[V][\ell\tau\tau\ell]$ involving the $\tau$, the semileptonic tensor operators of Eq.~\eqref{eq:a2l2q}, which  give a non-perturbative contribution to the dipole moments, and the semileptonic charm-tensor operator of Eq.~\eqref{eq:CharmLoop}.

Assuming only a single Wilson coefficient to be non-zero, the needed values of the Wilson coefficients to explain $\Delta a_\mu$ or $\Delta a_e$ individually are collected in the upper panel of Table~\ref{tab:LEFTWCsamuae}. For $\Delta a_e$, we give two sets of numbers using the two values for $\alpha_\mathrm{QED}$ in Eq.~\eqref{eq:alphaQED}. Of course, an explicit BSM model can generate more than one non-zero coefficient, as discussed for a leptoquark model in Sec.~\ref{sec:Models}. In the case of $\lwc{ee}{RR}[S][e\mu \mu e]$ and $\lwc{ee}{LR}[V][e \mu \mu e]$, the symmetry relation Eq.~\eqref{eq:symmetry} implies that they contribute to both muon and electron anomalies, which require different values. The second panel of Table~\ref{tab:LEFTWCsamuae} shows a minimal combined scenario for these two parameters, where both are non-zero, and adjusted to explain the muon anomaly while largely avoiding effects in $\Delta a_e$, which leads to a very strong correlation. Five significant digits are shown to illustrate that a simultaneous explanation of the electron anomaly $\Delta a_e^\mathrm{Cs}$ or $\Delta a_e^\mathrm{Rb}$ requires a fine-tuning of the two coefficients.
\begin{table}[H]
  \centering\small
  \renewcommand{\arraystretch}{1.9}
\begin{tabular}{lcrrr}
  	\toprule\\[-1.2cm]
  	coefficient & \text{units} & \multicolumn{1}{c}{$\ell=\mu$} & \multicolumn{1}{c}{$\ell=e$, Cs input} & \multicolumn{1}{c}{$\ell=e$, Rb input} \\
  	\hline
	$\lwc{e\gamma}{}[][\ell\ell]$ &												GeV${}^{-1}$ &			$\phantom{+}[1.4,2.3] \times10^{-9}$ &				$-[8.2,20]\times10^{-11}$					&	$\phantom{+}[2.8,12]\times10^{-11}$				\\
	$\lwc{ee}{LR}[V][\ell ee\ell]$ &												GeV${}^{-2}$  &		$\color{red}-[1.4,2.3]\times10^{-3}$&					$\phantom{+}[7.9,19]\times10^{-5}$			&	$-[2.7,12]\times10^{-5}$						\\
	$\lwc{ee}{LR}[V][\ell\mu\mu\ell]$  &											GeV${}^{-2}$ &			$-[6.8,11]\times10^{-6}$&							$\color{red}\phantom{+}[3.8,9.1]\times10^{-7}$	&	$\color{red}-[1.3,5.7]\times10^{-7}$				\\
	$\lwc{ee}{LR}[V][\ell\tau\tau\ell]$ &											GeV${}^{-2}$ &			$-[4.0,6.5]\times10^{-7}$&							$\phantom{+}[2.3,5.4]\times10^{-8}$			&	$-[7.8,34]\times10^{-9}$						\\
	$\lwc{ee}{RR}[S][\ell ee\ell]$ &												GeV${}^{-2}$ &			$\color{blue}\phantom{+}[8.5,14]\times10^{-5}$&		$-[4.8,11]\times10^{-6}$					&	$\phantom{+}[1.6,7.1]\times10^{-6}$				\\
	$\lwc{ee}{RR}[S][\ell\mu\mu\ell]$ &											GeV${}^{-2}$ &			$\phantom{+}[1.2,1.9]\times10^{-6}$&				$\color{blue}-[6.5,15]\times10^{-8}$			&	$\color{blue}\phantom{+}[2.2,9.7]\times10^{-8}$		\\
	$\lwc{ee}{RR}[S][\ell\tau\tau\ell]$ &											GeV${}^{-2}$ &			$\phantom{+}[1.7,2.8]\times10^{-6}$&				$-[9.6,23]\times10^{-8}$					&	$\phantom{+}[3.3,14]\times10^{-8}$				\\
	$c_T\lwc{eu}{RR}[T][\ell\ell uu]$  &											GeV${}^{-2}$ &			$-[4.7,7.5]\times10^{-7}$&							$\phantom{+}[2.6,6.2]\times10^{-8}$			&	$-[9.0,39]\times10^{-9}$						\\
	$c_T\lwc{ed}{RR}[T][\ell\ell dd]$, $c_T\lwc{ed}{RR}[T][\ell\ell ss]$  &					GeV${}^{-2}$ &			$\phantom{+}[9.3,15]\times10^{-7}$&					$-[5.2,12]\times10^{-8}$					&	$\phantom{+}[1.8,7.8]\times10^{-8}$				\\
	$c_T^{(c)}\lwc{eu}{RR}[T][\ell\ell cc]$  &										GeV${}^{-2}$ &			$\phantom{+}[7.1,11]\times10^{-8}$&					$-[4.0,9.5]\times10^{-9}$					&	$\phantom{+}[1.4,6.0]\times10^{-9}$				\\
	$c_S\lwc{eu}{RR}[S][\ell\ell uu]$  &											GeV${}^{-2}$ &			$\phantom{+}[1.2,1.9]\times10^{-3}$&				$-[6.5,16]\times10^{-5}$					&	$\phantom{+}[2.3,9.8]\times10^{-5}$				\\
	$c_S\lwc{ed}{RR}[S][\ell\ell dd]$, $c_S\lwc{ed}{RR}[S][\ell\ell ss]$  &					GeV${}^{-2}$ &			$\phantom{+}[4.7,7.6]\times10^{-3}$&				$-[2.6,6.2]\times10^{-4}$					&	$\phantom{+}[9.1,39]\times10^{-5}$				\\
	$c_S^{(c)}\lwc{eu}{RR}[S][\ell\ell cc]$  &										GeV${}^{-2}$ &			$-[1.8,2.9]\times10^{-4}$&							$\phantom{+}[1.0,2.4]\times10^{-5}$			&	$-[3.5,15]\times10^{-6}$						\\
	$\widetilde c_S\lwc{eu}{RL}[S][\ell\ell uu]$  &									GeV${}^{-2}$ &			$\phantom{+}[1.2,1.9]\times10^{-3}$&				$-[6.5,16]\times10^{-5}$					&	$\phantom{+}[2.3,9.8]\times10^{-5}$				\\
	$\widetilde c_S\lwc{ed}{RL}[S][\ell\ell dd]$, $\widetilde c_S\lwc{ed}{RL}[S][\ell\ell ss]$  &	GeV${}^{-2}$ &			$\phantom{+}[4.7,7.6]\times10^{-3}$&				$-[2.6,6.2]\times10^{-4}$					&	$\phantom{+}[9.1,39]\times10^{-5}$				\\
	$\widetilde c_S^{(c)}\lwc{eu}{RL}[S][\ell\ell cc]$  &								GeV${}^{-2}$ &			$-[1.8,2.9]\times10^{-4}$&							$\phantom{+}[1.0,2.4]\times10^{-5}$			&	$-[3.5,15]\times10^{-6}$						\\[0.15cm]
		\midrule
	$\lwc{ee}{LR}[V][e\mu\mu e]$ &												GeV${}^{-2}$  &		&											$1.0133\times10^{-3}$					&	$1.0117\times10^{-3}$						\\
	$\lwc{ee}{RR}[S][e\mu\mu e]$ &											GeV${}^{-2}$ &			&											$1.7218\times10^{-4}$					&	$1.7208\times10^{-4}$						\\[0.15cm]
	\bottomrule
  \end{tabular}
  \caption{Upper panel: size of single non-vanishing LEFT Wilson coefficients at $\mu=2\GeV$ needed to explain the anomalies $\Delta a_\mu$ or $\Delta a_e$ individually. The ranges correspond to $\pm1\sigma$ in Eqs.~\eqref{eq:MuonDiscrepancy} and~\eqref{eq:ElectronDiscrepancy}.
The red entries are the same coefficient $\lwc{ee}{LR}[V][e \mu \mu e]=\lwc{ee}{LR}[V][\mu e e\mu ]$ and the blue entries are the same coefficient $\lwc{ee}{RR}[S][e\mu \mu e]= \lwc{ee}{RR}[S][\mu e e \mu] $, from Eq.~\eqref{eq:symmetry}.
An explanation of the muon anomaly that avoids huge effects in $\Delta a_e$ is possible with the combination in the lower panel. A simultaneous explanation of $\Delta a_e$ requires a fine-tuning of the two coefficients.}
  \label{tab:LEFTWCsamuae}
\end{table}

The Wilson coefficients that enter Eq.~\eqref{eq:amuLEFTnum} are only weakly constrained. The  LHC measurements of $pp\to \mu^+\mu^-$ give a constraint on the semileptonic operator  $\big|  \lwc{eu}{RR}[T][\mu\mu uu] \big| \leq 9.3\, {\rm TeV}^{-2}$~\cite{Falkowski:2017pss}. However, this limit is only valid if  BSM physics arises well above the electroweak scale, $\Lambda \gg \sqrt{s}\sim$ TeV, an assumption that does not necessarily hold in the LEFT. A comparable constraint can be derived if one assumes that $\lwc{eu}{RR}[T][\mu\mu uu]$ is induced by the SMEFT Wilson coefficient $\cwc{\ell equ}{(3)}[][\mu\mu uu]$. In this case $SU(2)_L$ invariance implies that the LEFT operator is related to a charged-current interaction that can be probed by $\pi\to \mu\nu$~\cite{Falkowski:2017pss}.
Similarly, constraints on four-lepton operators can be derived if one assumes that $\lwc{ee}{LR}[V][\mu \tau \tau \mu ]$ originates from the SMEFT operator $\cwc{\!\!\!\ell e}{}[][\!\!\!\mu \tau \tau \mu]$. In this case, $\lwc{ee}{LR}[V][\mu \tau \tau \mu ]$ is related to a charged-current interaction that can be constrained by $\tau\to \mu\nu_\tau\bar \nu_\mu$, leading to $\big| \lwc{ee}{LR}[V][\mu \tau \tau \mu] \big| \leq 12 \,{\rm TeV}^{-2}$~\cite{Stahl:1999ui,Falkowski:2015krw,Zyla:2020zbs}. For both operators these bounds are not stringent enough to exclude the values collected in Table~\ref{tab:LEFTWCsamuae}. In addition, it should be noted that neither of the above constraints hold in scenarios where the SMEFT approach does not apply, as would be the case when BSM physics appears at or below the electroweak scale.

The expressions for the magnetic moments can be used to obtain the EDMs of the leptons, $d_\ell$, by multiplying Eq.~\eqref{eq:amuLEFTnum} by ${\mathsf{q}_e e_\mathrm{QED}}/{(2 m_\ell)} $ and replacing $\Re$ by $\Im$. Combined with the experimental bounds in Eq.~\eqref{eq:electric}, this allows one to constrain the imaginary parts of the Wilson coefficients.
These bounds can be determined by multiplying the middle of the ranges given in the $\ell=\mu$ and $\ell=e, \text{Cs}$ columns in Table~\ref{tab:LEFTWCsamuae} by the combination
$ 2m_\ell d_\ell^\mathrm{exp}/(e_\mathrm{QED} |\Delta a_\ell|) =  \{ (640.0)_\mu , (6.5 \times 10^{-7})_e \}$ for the muon and electron, respectively, with $\Delta a_\ell$ and $d_\ell^{\rm exp}$ given by Eqs.~\eqref{eq:MuonDiscrepancy}, \eqref{eq:ElectronDiscrepancy}, and \eqref{eq:electric}. From the values of these rescaling factors one sees that $d_\mu^{\rm exp}$ allows the imaginary parts of the Wilson coefficients to be a factor of $\sim 6 \times 10^2$ larger than the size of the real parts required to explain $\Delta a_\mu$. Instead, for the case of the electron, the imaginary parts are constrained to be roughly six orders of magnitude smaller than the size of the real parts needed to explain $\Delta a_e$, leading to limits $\sim 10^{-10}-10^{-15} \,{\rm GeV}^{-2}$ on the imaginary parts of the coefficients of the four-fermion operators.\footnote{Of course, the imaginary parts of the flavor-diagonal $V,LR$ operators vanish due to the symmetry relation of Eq.~\eqref{eq:symmetry}.}

Although simple rescaling can be used to obtain limits on most of the $CP$-odd Wilson coefficients, it is not appropriate in all cases due to the fact that currently the most stringent limit on $d_e$ results from measurements on the paramagnetic molecule ThO. $CP$-odd effects in this molecule are not only induced by $d_e$, but also by electron-nucleon interactions of the form $\bar NN\, \bar e i\gamma_5 e$~\cite{doi:10.1063/1.4968597,Skripnikov,Skripnikov:2013}, generated by the semileptonic scalar operators in Eq.~\eqref{eq:SMFourFermiA} that contain light quarks. Similarly, the stringent limit on the EDM of the diamagnetic atom $d_\mathrm{Hg}$~\cite{Graner:2016ses,Griffith:2009zz} can be used to probe spin-dependent nucleon-electron interactions that are induced by the semileptonic tensor operators in Eq.~\eqref{eq:SMFourFermiA} involving light quarks \cite{Fleig:2018bsf}. Following Refs.~\cite{Dekens:2018pbu,Dekens:2018bci}, these effects lead to constraints of the order of $\sim 10^{-15} \,{\rm GeV}^{-2}$ on the semileptonic scalar and $\sim 10^{-14} \,{\rm GeV}^{-2}$ on the semileptonic tensor operators.

The discrepancies $\Delta a_\mu$ and $\Delta a_e$ can be explained simultaneously within the LEFT by choosing suitable linear combinations of the Wilson coefficients in Table~\ref{tab:LEFTWCsamuae}. This result is not surprising since LEFT contains a dimension-five dipole operator that contributes to the dipole moments. The values in Table~\ref{tab:LEFTWCsamuae} also provide an estimate of the scale at which BSM physics must occur to be able to explain the dipole anomalies, based on the low-energy operator generated in the model. Naive dimensional analysis (NDA) scaling~\cite{Manohar:1983md,Gavela:2016bzc} of the operators Eq.~\eqref{eq:SMFourFermi} gives $L_{5} \lesssim 4\pi/\Lambda_\mathrm{NDA}$ for the dimension-five dipole operators, and  $L_6 \lesssim (4\pi/\Lambda_\mathrm{NDA})^2$ for the dimension-six operators, where $\Lambda_\mathrm{NDA}$ is the scale of new physics as given by the derivative expansion, i.e.\ the $p/\Lambda_\mathrm{NDA}$ expansion. These inequalities
can be interpreted as bounds on the coefficients if the theory is strongly coupled, i.e.\ at the unitarity bound. The Wilson coefficients can have additional perturbative suppression factors if some of the interactions are weakly coupled. For example, if the dipole operator has a coupling constant $g$, the inequality becomes $L_{5} \lesssim g / \Lambda_\mathrm{NDA}$. If the photon is weakly coupled in the BSM theory, one expects a coupling constant $g=e$ in the dipole-operator coefficient. Similarly, if the four-fermion operators have a factor of $g^2$, as occurs in the low-energy operators generated by weak gauge-boson exchange, then $L_6 \lesssim g^2/ \Lambda_\mathrm{NDA}^2$. The SM electroweak coupling $g_2 \sim 0.6$, so in the following discussion, we will use $L_5 \sim 1/\Lambda$ and $L_6 \sim 1/\Lambda^2$ to estimate the scale of new physics. One can get the corresponding values of $\Lambda_\mathrm{NDA}$ for a strongly coupled theory by multiplying $\Lambda$ by $4\pi$.

For $\Delta a_e^\mathrm{Cs}$, the values in Table~\ref{tab:LEFTWCsamuae} mostly correspond to rather high BSM scales: for the dimension-five dipole operator, the scale is $\Lambda \sim 1/ \abs{  \lwc{}{}[][] } \sim 10^7\TeV$, while in the case of most four-fermion operators, the scale $\Lambda \sim 1/\sqrt{ \abs{ \lwc{}{}[][] }}$ is still about $1-5\TeV$ or even $12\TeV$ for $\lwc{eu}{RR}[T][eecc]$. The two coefficients $\lwc{ee}{LR}[V][eeee]$ and $\lwc{ee}{RR}[S][eeee]$ would need to arise at the lower scales $\Lambda\sim 90$ GeV and $\Lambda\sim 350$ GeV, respectively. The scalar semileptonic operators $\lwc{ed}{RR}[S][]$, $\lwc{eu}{RR}[S][]$, $\lwc{ed}{RL}[S][]$, $\lwc{eu}{RL}[S][]$ also need low scales in the $50-200\GeV$ range.

In the case of $\Delta a_\mu$, the corresponding BSM scales are generally lower, $10^6\TeV$ for the dipole operator, and ranging between $300\GeV$ to about a TeV for most of the four-fermion operators. The exceptions are again $\lwc{eu}{RR}[T][\mu\mu cc]$, pointing to $3\TeV$, and the couplings to the first generation, $\lwc{ee}{LR}[V][e \mu \mu e]$ and $\lwc{ee}{RR}[S][e \mu \mu e]$, which would have to be induced at  $\Lambda\sim 20$ GeV and $\Lambda \sim 90$ GeV, respectively, to explain $\Delta a_\mu$. The semileptonic operators $\lwc{ed}{RR}[S][]$, $\lwc{eu}{RR}[S][]$, $\lwc{ed}{RL}[S][]$, $\lwc{eu}{RL}[S][]$ require low scales in the $10-50\GeV$ range.

While the scale for the dipole operators is high, in most models, the dipole coefficients are effectively dimension six, $L \sim v/\Lambda^2$, so that
$\Delta a_e^\mathrm{Cs}$ requires a dipole scale $\Lambda \sim 10^3\TeV$ and $\Delta a_\mu$ a dipole scale $\Lambda \sim 360\TeV$. If BSM models induce dipole operators with a loop suppression of $\sim \alpha/(4\pi)$, the scale is lowered even further, to $\sim 30\TeV$ and $\sim 9\TeV$, respectively. If the four-fermion operators are generated in BSM models with small coupling constants, then the scale at which they are generated will be lower than the values given here.

The values presented in this section are for couplings renormalized at the scale $\mu=2\GeV$. We include RG running effects and operator mixing between $\Lambda$ and $2\GeV$ in the next sections.


\section{Running effects below the electroweak scale}
\label{sec:LEFT}

In the previous section, we investigated the LEFT contributions to the anomalous magnetic moments using parameters at the low scale $\mu=2\GeV$. The LEFT is valid up to the scale of BSM physics, or the electroweak scale $\mu_W$, whichever is lower. We can run the LEFT parameters from $\mu=2\GeV$ to a higher scale, to investigate the effects of RG evolution, which is due to QCD and QED. To illustrate this running effect and its implications on the anomalous magnetic moment, we will scale Eq.~\eqref{eq:amuLEFTnum} from $2\GeV$ up to $\mu=60\GeV$ in the LEFT. Taking one-loop QCD and QED anomalous dimensions \cite{Jenkins:2017dyc,Aebischer:2017gaw,Gonzalez-Alonso:2017iyc} into account, one finds for the electron or muon magnetic moment, with terms in descending order of size of their coefficients,
\begin{align}\label{eq:amu60}
& \Delta a_{\ell}^{60\GeV} =\frac{m_\ell}{m_\mu}\Re\Bigg[
{2.2_\mu \atop 2.1_e} \times 10^{-2}  \tlwc{e\gamma}{}[][\ell \ell ]
 - {5.3_\mu \atop 5.1_e} \times 10^{-5}  \tlwc{ed}{RR}[T][\ell\ell bb ]  \nn
& + \left({3.5_\mu \atop 3.4_e} +0.65 c_T^{(c)} \right) \times 10^{-5} \tlwc{eu}{RR}[T][\ell\ell cc ]
 +{ 9.0_\mu \atop 8.7_e} \times 10^{-6}  \tlwc{ee}{RR}[S][\ell \tau \tau \ell ]
- 1.4 \times 10^{-6}  \tlwc{ee}{LR}[V][\ell \tau \tau \ell ] \nn
& + { 9.8_\mu \atop 9.6_e } \times 10^{-7}  \tlwc{ee}{RR}[S][\ell \mu\mu \ell ]
 - \left(10 c_T - {0.64_\mu \atop 0.62_e} \right) \times 10^{-7}   \tlwc{eu}{RR}[T][\ell\ell uu ]
 + \left(5.0  c_T - {14_\mu \atop 13_e} \right) \times 10^{-7}  \tlwc{ed}{RR}[T][\ell\ell ss ] \nn
& + \left(5.0 c_T - {0.70_\mu \atop 0.67_e}\right) \times 10^{-7}  \tlwc{ed}{RR}[T][\ell\ell dd ]
 - { 1.6_\mu \atop 1.5_e} \times 10^{-7}  \tlwc{ee}{RR}[S][\ell\ell \tau \tau ] \nn
& - \left({5.9_\mu \atop 5.7_e} +2.3 c_T^{(c)}+0.45 c_S^{(c)} \right) \times 10^{-8} \tlwc{eu}{RR}[S][\ell \ell c c]
 - {8.0_\mu \atop 8.1_e } \times 10^{-8} \tlwc{ee}{LR}[V][\ell \mu \mu \ell ]
 - {3.3_\mu \atop 3.2_e} \times 10^{-8}  \tlwc{ed}{RR}[S][\ell\ell b b ] \nn
& -2.4  \times 10^{-8} \tlwc{ee}{RR}[S][\ell\ell \mu \mu ]
 + {8.8_\mu \atop 8.4_e} \times 10^{-9}  \tlwc{ee}{RR}[S][\ell e e \ell ]
 - 4.5 \times 10^{-9} \widetilde c_S^{(c)} \tlwc{eu}{RL}[S][\ell \ell c c]
+ 3.5 \times 10^{-9} c_T  \tlwc{eu}{RR}[S][\ell \ell uu ]  \nn
& - {1.2_\mu \atop 1.1_e} \times 10^{-9}  \tlwc{ed}{RR}[S][\ell \ell ss ]
\Bigg] \, ,
\end{align}
where we have used the dimensionless constants
\begin{align}\label{eq:scale}
 \tlwc{i}{}[][]\equiv \Lambda^{-d_i}\lwc{i}{}[][](\mu=\Lambda),
 \end{align}
with $d_i$ the dimension of $\lwc{i}{}[][]$ and $\Lambda = 60\GeV$, and we have dropped terms with coefficients smaller than $10^{-9}$. If the BSM scale is $60\GeV$, the dimensionless constants $\tilde L$ are expected to be at most of order unity, and can be smaller if there are suppressions due to small coupling constants.
Note that the contribution of $\lwc{ee}{RR}[S][\ell \ell \tau \tau]$ is generated by first running into the $\lwc{ee}{RR}[S][\ell \tau \tau \ell]$, which subsequently induces the dipole operator, effectively making it a two-loop effect. Similar two-loop mixing effects occur for the semileptonic scalar operators. Apart from the effect of the scalar operators, the RG resummation beyond one loop typically induces corrections of only a few percent, reaching $6.3\%$ in the case of the $b$-quark tensor operator $\lwc{ed}{RR}[T][\ell\ell bb]$. The non-perturbative contributions at low energies, which depend on the parameters $c_T$, $c_S$ and $\widetilde c_S$, are important for the light-quark operators.

As the scale of new physics  increases, the size of $L_i$ decreases, and eventually the contribution from some operators becomes too small to explain the $g-2$ discrepancy. At the same time, RG evolution and operator mixing implies that new operators can contribute via mixing into those present in Eq.~\eqref{eq:amuLEFTnum}.
All Wilson coefficients present at $\mu=2\GeV$ still contribute at $\Lambda=60\GeV$, although $\tlwc{ee}{LR}[V][\ell ee\ell ]$ has been dropped since its coefficient is a bit smaller than $10^{-9}$. Furthermore, two additional semileptonic $b$-quark operators, as well as additional leptonic scalar operators, contribute to $a_\ell$, when RG effects are taken into account:
\begin{equation}\label{eq:60new}
\lwc{ed}{RR}[T][\ell \ell b b ]\,,\quad\lwc{ed}{RR}[S][\ell \ell bb]\,,\quad\lwc{ee}{RR}[S][\ell \ell \tau \tau ]\,,\text{ and, in the case of $a_e$, } \lwc{ee}{RR}[S][ee\mu\mu] \, .
\end{equation}
The Wilson coefficient $\lwc{ed}{RR}[T][\ell \ell b b ]$ in Eq.~\eqref{eq:60new} contributes to $a_\ell$ by mixing into the dipole operator via QED penguin diagrams. Its rather large contribution results from  chiral enhancement: the chirality change inside the $b$-quark loop provides a mass factor $m_b$, instead of $m_\ell$.

For the example of $a_\mu$, we again convert the required value of $\widetilde L$ to a naive BSM scale $\Lambda$ (without taking into account possible additional BSM couplings that could lower the scales). The Wilson coefficients that correspond to the highest probed scales are shown in Fig.~\ref{fig:barplot}, reaching almost $8\TeV$ for the case of the tensor-operator coefficient $\lwc{eu}{RR}[T][\mu\mu cc]$ and about $9\TeV$ for the case of $\lwc{ed}{RR}[T][\mu \mu b b]$. Such high scales motivate an analysis within the SMEFT instead of the LEFT, which we will consider in the next section.\footnote{Note that $\lwc{ed}{RR}[T][]$ is not generated in the SMEFT up to dimension six and one loop~\cite{Dekens:2019ept}.}
\begin{figure}[tbp]
	\centering
\begin{tikzpicture}
\definecolor{myorange}{HTML}{fe7f10}
\definecolor{myblue}{HTML}{1f77b4}
\begin{axis} [ybar=0pt,
	height=8cm,width=12cm, bar width = 14pt,
	ymode = log, log origin=infty,
	ymin = 8.e-1, ymax = 2.e3,
	xmin = 0.15, xmax = 8.75,
	axis lines = none
]
\pgfplotsset{
	legend image code/.code={
	\draw [#1] (0cm,-0.1cm) rectangle (0.6cm,0.1cm);
	},
}
\addplot[fill = myorange, draw = myorange] coordinates {
	(1,	3.57e2)
	(2,	8.71e0)
	(3,	7.69e0)
	(4,	3.59e0)
	(5,	1.65e0)
	(6,	1.39e0)
	(7,	1.24e0)
	(8,	1.17e0)
	};
\addplot[fill = myblue, draw = myblue] coordinates {
	(1,	1.30e3)
	(2,	3.18e1)
	(3,	2.82e1)
	(4,	1.31e1)
	(5,	6.04e0)
	(6,	5.18e0)
	(7,	4.59e0)
	(8,	4.36e0)
	};
\legend {$\Delta a_\mu$, $\Delta a_e^\mathrm{Cs}$};
\end{axis}
\begin{axis} [ybar,
	height=8cm,width=12cm,
	ylabel = {$\Lambda$ [TeV]},
	axis line style={line width=0.5pt},
	ymode = log, log origin=infty,
	ymin = 8.e-1, ymax = 2.e3,
	xmin = 0.15, xmax = 8.75,
	tick style={line width=0.75pt,draw=black},
	major tick length=8pt,
	minor tick length=4pt,
	xtick=\empty,
	xtick pos=left, ytick pos=left,
	extra x ticks={1,2,3,4,5,6,7,8},
	extra x tick style={major tick length=4pt},
	extra x tick labels={
    		$\lwc{e\gamma}{}[][\ell\ell]$,
    		$\lwc{ed}{RR}[T][\ell\ell bb]$,
    		$\lwc{eu}{RR}[T][\ell\ell cc]$,
    		$\lwc{ee}{RR}[S][\ell\tau\tau\ell]$,
    		$\lwc{ed}{RR}[T][\ell\ell ss]$,
    		$\lwc{ee}{LR}[V][\ell\tau\tau\ell]$,
    		$\lwc{eu}{RR}[T][\ell\ell uu]$,
    		$\lwc{ee}{RR}[S][\ell\mu\mu\ell]$
	},
]
\end{axis}
\end{tikzpicture}
\caption{Eight largest scales probed by the anomalous magnetic moments of the muon (orange) and the electron (blue) of the LEFT Wilson coefficients in Eq.~\eqref{eq:amu60}, illustrated for the choices $c_T^{(c)}=-c_T=1$ and $\Delta a_e^\mathrm{Cs}$ based on the Cesium input for $\alpha_\mathrm{QED}$. For the dipole-operator coefficients, we assume the SMEFT scaling $\lwc{e\gamma}{} \sim v / \Lambda^2$.}
\label{fig:barplot}
\end{figure}
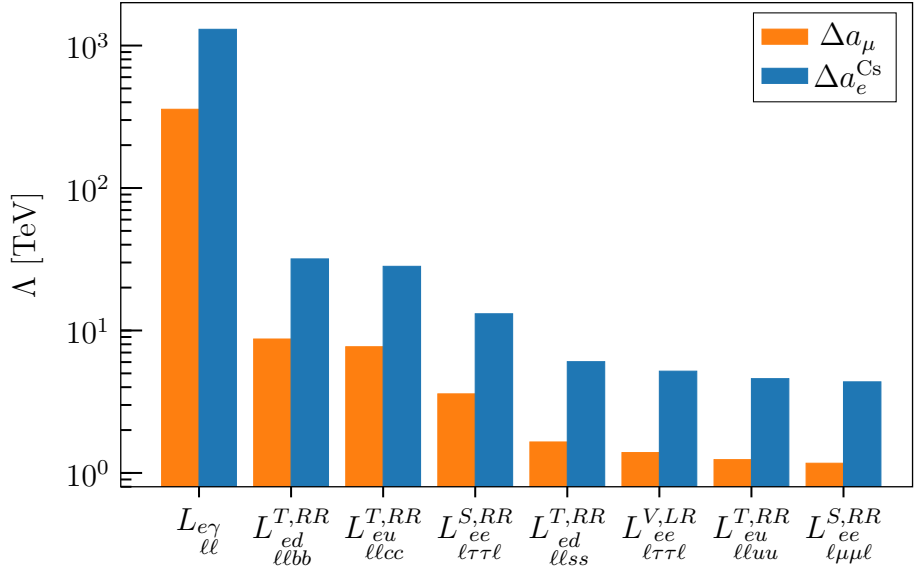

With an increasing scale of new physics, more correlations between different observables and additional constraints arise and one might wonder if the dipole operators themselves can be constrained through their mixing effects. Since mixing only is possible with lower-dimensional operators, one could consider the effects on the running of the electromagnetic coupling. Dipole insertions in a vacuum-polarization diagram, as shown in Fig.~\ref{fig:RunningAlpha}, induce the running~\cite{Jenkins:2017dyc}
\begin{align}
	16\pi^2 \mu\frac{d}{d\mu} e = -16 e^2 \q_e m_\ell\, \Re \lwc{e\gamma}{}[][\ell\ell] \, .
\end{align}
The dominant contribution is from the $\tau$ dipole moment because of the mass dependence. Defining
\begin{align}
	\alpha^{-1}(\mu=M_Z) = \alpha^{-1}(\mu=0)( 1 - \Delta\alpha )
\end{align}
we find
\begin{align}
	\Delta\alpha\big|_\mathrm{dipoles} = 3.8 \times 10^{-6} \tlwc{e\gamma}{}[][ee] + 4.6 \times 10^{-4} \tlwc{e\gamma}{}[][\mu\mu] + 4.6 \times 10^{-3} \tlwc{e\gamma}{}[][\tau\tau] \, ,
\end{align}
with dimensionless dipole-operator coefficients defined as in Eq.~\eqref{eq:scale}, but for $\Lambda=M_Z$. Determinations of the QED coupling at the weak scale through global electroweak fits are largely compatible with input on the hadronic running from $e^+e^-$ data~\cite{Davier:2019can,Keshavarzi:2019abf}. The electroweak fits lead to a precision on $\Delta\alpha$ at the weak scale of $\sim 39 \times 10^{-5}$~\cite{Haller:2018nnx}, which constrains the rescaled $\tau$-dipole-operator coefficient at the level of $8.4\times10^{-2}$. This corresponds to a tree-level contribution to $a_\tau$ of
\begin{align}
	|\Delta a_\tau| < 2.1\times 10^{-2} \, ,
\end{align}
which is comparable to the constraint on $a_\tau$ itself~\cite{Zyla:2020zbs}. However, this bound only applies if the $\tau$-dipole operator is generated at or above the electroweak scale. If it is loop induced, the suppression factor pushes the naive scale well below the weak scale.

The constraints from the running of $\alpha$ have mainly been studied in the context of the HVP contribution to $a_\mu$~\cite{Passera:2008jk,Crivellin:2020zul,Keshavarzi:2020bfy,Malaescu:2020zuc,Colangelo:2020lcg}. In Ref.~\cite{deRafael:2020uif}, an EFT language was employed, parameterizing changes in the HVP function in terms of a dimension-six operator $\p^\lambda F^{\mu\nu} \p_\lambda F_{\mu\nu}$. In the LEFT basis, this operator is replaced via the equations of motion by a set of current-current operators including $\Op{ee}{LR}[V][]$. This might seem puzzling at first sight, as after the basis change the four-fermion operators do not contribute to the photon polarization function, whereas the equation-of-motion operator did. However,
to relate the effective running QED coupling to observables, it should not be defined as usual in terms of the basis-dependent photon 1PI function, but rather through four-fermion processes, which include the contact contributions of four-fermion operators. As before, the EFT approach only applies if the scale of the operator lies at or above the electroweak scale. In Ref.~\cite{deRafael:2020uif}, changes in the HVP function were related to a scale of a few GeV, so that no model-independent connection between the HVP contribution to $a_\mu$ and $\alpha$ at the weak scale could be established.
\begin{figure}[tbp]
	\centering
	\scriptsize
	\begin{fmfgraph*}(80,40)
		\fmfleft{l1} \fmfright{r1}
		\fmf{photon,label=$\gamma$,tension=3}{l1,v1}
		\fmf{photon,label=$\gamma$,tension=3}{v2,r1}
		\fmf{quark,right,label=$\ell$,label.side=right}{v1,v2,v1}
		\fmfblob{3mm}{v1}
	\end{fmfgraph*}\\[0.25cm]
	\caption{Mixing between dipole operators and the QED charge through the vacuum polarization graph.}
	\label{fig:RunningAlpha}
\end{figure}
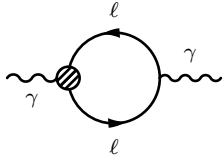


\section{SMEFT interpretation}
\label{sec:SMEFT}

Above the EW scale $\mu_W$, the LEFT is replaced by the SMEFT which consists of the SM plus higher dimension operators. The LEFT coefficients can be computed in terms of those in SMEFT by matching the two EFTs at the electroweak scale. The matching equations have been computed at tree level~\cite{Jenkins:2017jig,Liao:2020zyx} and at one loop~\cite{Aebischer:2015fzz,Dekens:2019ept,Hurth:2019ula}, and we include the one-loop effects. The heavy SM particles, i.e., the top quark, $W$, $Z$, and Higgs bosons, are integrated out at one common scale $\mu_W$, which we set equal to the \msbar{} top mass, $\mu_W = m_t = 162.5\GeV$~\cite{Zyla:2020zbs}. As an input in the matching equations, we use the on-shell $W$, $Z$ boson masses and $\alpha_\mathrm{QED}$ evolved from the scale $\mu=M_Z$ to $\mu=m_t$ at one loop. The exact choice of  scheme changes the numerical results by a few percent. For simplicity, we neglect additional dimension-6 corrections that appear when expressing the SM electroweak contribution to $a_\ell$ in terms of the Fermi constant determined from muon decay, as the discrepancy $\Delta a_\mu$ is larger than the SM electroweak contribution Eq.~\eqref{eq:amuEWTwoLoops}.

The SMEFT results will be given at two scales, $\Lambda=250\GeV$ and $\Lambda=10\TeV$, to illustrate the contributions to the dipole moments. Within the SMEFT, we evolve the coefficients from the matching scale $\mu_W$ to the scale $\Lambda$ using the one-loop RG evolution~\cite{Jenkins:2013zja,Jenkins:2013wua,Alonso:2013hga}. In terms of SMEFT Wilson coefficients renormalized at $\Lambda=250\GeV$, this results in the following expression for the magnetic moments:
\begin{align}\label{eq:amuSMEFT250}
\Delta a_{\ell}^{250\GeV} &=\frac{m_\ell}{m_\mu}\Re\Bigg[{2.9_\mu \atop 2.8_e} \times 10^{-3}\tcwc{eB}{}[][\ell\ell]-{1.6_\mu \atop 1.5_e}\times 10^{-3}\tcwc{eW}{}[][\ell\ell]\nn
&\quad -{4.3_\mu \atop 4.1_e} \times 10^{-5}\tcwc{\ell equ}{}[(3)][\ell\ell33]-\left(2.6 +0.37 c_T^{(c)}\right)\times 10^{-6}\tcwc{\ell equ}{}[(3)][\ell\ell22] \nn
&\quad  -7.9\times 10^{-8}\tcwc{\ell e}{}[][\ell33\ell] + \left(5.7c_T-{0.49_\mu \atop 0.48_e}\right)\times 10^{-8}\tcwc{\ell equ}{}[(3)][\ell\ell11]  +1.4\times 10^{-8}\tcwc{\ell equ}{}[(1)][\ell\ell33] \nn
 &\quad +\left({10_\mu \atop 9.8_e}+2.5 c_T^{(c)}\right)\times 10^{-9}\tcwc{\ell equ}{}[(1)][\ell\ell22]  -{4.6_\mu \atop 4.7_e} \times 10^{-9}\tcwc{\ell e}{}[][\ell22\ell] \nn
\nn[-0.5cm]
&\quad + \frac{m_\ell}{m_\mu} \bigg\{ {2.5_\mu \atop 2.4_e} \times 10^{-8} \left( \tcwc{HWB}{}[][] + i \tcwc{H \widetilde W B}{}[][] \right) -{1.8_\mu \atop 1.7_e} \times 10^{-8} \left( \tcwc{HB}{}[][] + i \tcwc{H \widetilde B}{}[][] \right) \nn
&\quad -{6.0_\mu \atop 5.7_e} \times 10^{-9} \left( \tcwc{HW}{}[][] + i \tcwc{H \widetilde W}{}[][] \right) + 3.8\times 10^{-9} \tcwc{He}{}[][\ell\ell]  - {3.7_\mu \atop 3.6_e} \times 10^{-9} \tcwc{Hl}{}[(1)][\ell\ell] \nn
&\quad + {3.6_\mu \atop 3.3_e} \times 10^{-9} \tcwc{Hl}{}[(3)][\ell\ell] +{1.8_\mu \atop 1.7_e} \times 10^{-9} \tcwc{HD}{}[][]  + {2.1_\mu \atop 2.0_e} \times 10^{-9} \tcwc{W}{}[][] \nn
&\quad + 1.1\times 10^{-9} i \tcwc{\widetilde W}{}[][]
\bigg\} \Bigg] \, ,
\end{align}
where we used the mass basis for the up-type quarks and charged leptons and only contributions up to $10^{-9}$ have been retained. The $C_W$ and $ C_{\widetilde W}$ contributions violate holomorphy~\cite{Alonso:2014rga} due to the one-loop matching contribution, which depends on the scheme choice for $\mathcal Q_{\widetilde W}$, see the discussion in Ref.\ \cite{Dekens:2019ept} for details.
We have used the convention of Eq.~\eqref{eq:scale}, where couplings with a tilde are couplings scaled by powers of $\Lambda=250\GeV$ to make them dimensionless. The results for the dipole operators and $\cwc{\ell e}{}[][]$ are comparable to the findings in Ref.~\cite{Crivellin:2013hpa}. The terms from line five to eight in Eq.~\eqref{eq:amuSMEFT250} are shown for completeness but are subject to strong constraints. Using the python code \texttt{smelli}~\cite{Aebischer:2018iyb} and assuming only one operator is present at a time, these operators are excluded as explanations for the discrepancy: one finds that $\cwc{HWB}{}[]$, $\cwc{HB}{}[]$, $\cwc{HW}{}[]$, $\cwc{HD}{}[]$, and $\cwc{W}{}[]$ are strongly constrained by the signal strength of $h\to Z\gamma$ from gluon fusion production, $\mu_{gg}(h\to Z\gamma)$; $\cwc{He}{}[][22]$ is excluded by the $Z$-pole observable $R_\mu$, whereas $\cwc{H\ell}{}[(1)][22]$, $\cwc{H\ell}{}[(1)][11]$, $\cwc{He}{}[][11]$, and $\cwc{H\ell}{}[(3)][22]$ give too large contributions to $R(K^*)$ and $R(D^*)$, respectively. Finally, $\cwc{H\ell}{}[(3)][11]$ is constrained by a combination of meson decays.

In the end, very few SMEFT operators can explain the $a_\mu$ discrepancy; the electroweak dipoles $\cwc{eB}{}[][],  \cwc{eW}{}[][]$, and the four-fermion operators
$\cwc{\ell e}{}[][], \cwc{\ell e q u}{}[(3)][], \cwc{\ell e q u}{}[(1)][]$. The particle content of
possible tree-level models \cite{deBlas:2017xtg} generating the Wilson coefficients in Eq.~\eqref{eq:amuSMEFT250} that are not excluded phenomenologically are summarized in Table~\ref{tab:models}.

\begin{table}[tbp]
  \centering
  \begin{tabular}{clccccc}
  \toprule
    Spin & Rep. &
    $\mathcal{O}_{eB}$ &
    $\mathcal{O}_{eW}$ &
    $\mathcal{O}_{\ell e}$ &
    $\mathcal{O}^{(1)}_{\ell equ}$ &
    $\mathcal{O}^{(3)}_{\ell equ}$
    \\
    \midrule
    \multirow{3}{*}{0} & $\left({\bf 1},{\bf 2},1/2\right)$ &
    &&$\times$&$\times$&\\
    & $\left({\bf  3},{\bf 1},-1/3\right)$ &
    &&&$\times$&$\times$\\
    & $\left({\bf 3},{\bf 2},7/6\right)$ &
    &&&$\times$&$\times$\\
    \midrule
    \multirow{3}{*}{$\frac{1}{2}$} & $\left({\bf 1},{\bf 1},-1\right)$ &$\times$
    &&&&\\
    & $\left({\bf 1},{\bf 2},-1/2\right)$ &$\times$
    &$\times$&&&\\
    & $\left({\bf 1},{\bf 3},-1\right)$ &
    &$\times$&&&\\
\midrule
    \multirow{3}{*}{1} & $\left({\bf 1},{\bf 1},0\right)$ &
    &&$\times$&&\\
    & $\left({\bf 1},{\bf 2},1/2\right)$ &$\times$
    &$\times$&$\times$&&\\
    & $\left({\bf 1},{\bf 2},-3/2\right)$ &
    &&$\times$&&\\
    \bottomrule
  \end{tabular}
  \caption{Scalar, fermion, and vector representations that generate SMEFT operators at tree level~\cite{deBlas:2017xtg} relevant for $(g-2)$.
  }
  \label{tab:models}
\end{table}

As before, the EDMs of the leptons can be obtained from Eq.\ \eqref{eq:amuSMEFT250} by a simple rescaling and replacing $\Re$ by $\Im$. For example, to obtain $|d_\ell/d_\ell^{\rm exp}|$ the relevant factor is given by $e/(2m_\ell\,d^{\rm exp}_\ell) = \{(6.2\times 10^{5})_\mu ,\,  (1.8\times 10^{18})_e\}$. Thus, the experimental limit on $d_\mu$ allows for sizable imaginary parts in the Wilson coefficients $\tcwc{i}{}$ and only the dipole operators and the semileptonic tensors, $\tcwc{\ell equ}{}[(3)][2233]$ and $\tcwc{\ell equ}{}[(3)][2222]$, are constrained to be smaller than $1$.
The limits due to $d_e^{\rm exp}$ are much stronger, leading to nontrivial constraints even for the contributions that are suppressed by an additional factor of $m_e/m_\mu$.
Several of the constraints that can be obtained in this way were recently discussed in Refs.\ \cite{Pruna:2017tif,Panico:2018hal,Cirigliano:2019vfc}. In comparison, Ref.\ \cite{Panico:2018hal} is able to constrain additional operators by considering two-loop diagrams that induce next-to-leading-log contributions which were neglected in our analysis. On the other hand, we explicitly include the, sometimes sizable, non-perturbative contributions due to the semileptonic operators, which we parameterized by $c_{S,T}$ and $c_{S,T}^{(c)}$ and which were not considered in Ref.~\cite{Panico:2018hal}.

In a next step, we evolve the SMEFT Wilson coefficients to a higher scale of $\Lambda = 10\TeV$, finding for the magnetic moments:
\begin{align}\label{eq:amuSMEFT10TeV}
\Delta a_{\ell}^{10\TeV} =\frac{m_\ell}{m_\mu}\Re\Bigg[ & 1.7 \times 10^{-6}\tcwc{eB}{}[][\ell\ell]-{9.2_\mu \atop 8.9_e} \times 10^{-7}\tcwc{eW}{}[][\ell\ell]- {2.2_\mu \atop 2.1_e} \times 10^{-7}\tcwc{\ell equ}{}[(3)][\ell\ell33] \nn
	& - \left( {2.5_\mu \atop 2.4_e} + 0.22 c_T^{(c)} \right) \times 10^{-9}\tcwc{\ell equ}{}[(3)][\ell\ell22]
\Bigg] \, .
\end{align}
If the BSM scale is in the multi-TeV range, the only possible operators that provide viable explanations of the muon $g-2$ discrepancy are the electroweak dipole operators  $\cwc{eW}{}[][]$ and $\cwc{eB}{}[][]$, and the semileptonic tensor operators $\cwc{\ell equ}{(3)}[][]$.


\section{Leptoquarks}
\label{sec:Models}

The EFT analysis presented in the previous sections can be connected to specific models proposed to explain the muon anomaly. In this section, we review a particular example in the form of scalar leptoquark models, which can provide a NP contribution that is large enough to account for the discrepancy. The leptoquark models together with several other tree-level mediators generating the SMEFT Wilson coefficients are collected in Table~\ref{tab:models}.

Leptoquark models are prominent candidates to explain the  $a_\mu$ discrepancy (see, for example, Refs.~\cite{Dorsner:2016wpm,ColuccioLeskow:2016dox,Dorsner:2019itg,Crivellin:2020mjs,Crivellin:2020tsz} and references therein). The possible leptoquark models include $SU(2)_L$ singlet leptoquarks $S_{1}$ transforming as  $( {\bf \bar 3},{\bf 1},1/3)$ under $SU(3)_c \times SU(2)_L \times U(1)_Y$,
$SU(2)_L$ triplet leptoquarks $S_{3}$ transforming as $ ({\bf \bar 3},{\bf 3},1/3)$, or $SU(2)_L$ doublet leptoquarks $R_{2}$, transforming as $( {\bf  3},{\bf 2},7/6)$. Direct collider searches at the LHC constrain leptoquarks to be heavier than $1-1.5\TeV$~\cite{Aaboud:2019jcc,Camargo-Molina:2018cwu}. For the purposes of this paper, the inputs we need from the BSM theory are the leptoquark masses, and the Yukawa couplings of the leptoquarks to SM particles,
\begin{align}
\mathcal{L}_{S} &= y_{pr}^L\; ({q}^{T\alpha i }_p C  \ell^j_r ) \epsilon_{ij} S_{1\alpha} +y_{pr}^R\; ( {u}^{T \alpha}_p C e_{r} )  S_{1\alpha} + y^S_{pr}\; ( {q}^{T \alpha i }_p  C \ell^{j}_r )  \left(\epsilon \tau^I \right)_{ij}  S^I_{3 \alpha} \nn
&+y^{(u)}_{pr}\: (\bar{u}_{p \alpha} l_{r}^{j} ) \, \epsilon_{ij} R_{2}^{i \alpha} + y^{(q)}_{pr}\: ( \bar q^i_{p \alpha} e_r) \, R_{2}^{i \alpha} +\text{h.c.} \,,
\label{LY}
\end{align}
where $C = i \gamma^2 \gamma^0$ is the charge conjugation matrix,  $p,r$ are generation indices,  $i,j,I$ are weak $SU(2)_L$ indices, and $\alpha$ is a color index.
In these expressions, $q$ and $\ell$ are left-handed $SU(2)_L$ quark and lepton doublets, while $d$, $u$, and $e$ are right-handed $SU(2)_L$ singlet down-type, up-type,  and charged lepton fields.   The leptoquark Yukawa couplings $y^L_{pr} , y^R_{pr}$, $y^S_{pr}$, $y^{(u)}_{pr}$, and $y^{(q)}_{pr}$ are  arbitrary $3\times 3$ matrices in  flavor space.

At energy scales below the leptoquark mass, the leptoquarks can be integrated out,  generating four-fermion SMEFT operators with the tree-level coefficients
\begin{align}\label{eq:LQtree}
\cwc{\ell equ}{}[(1)][prst] &= \frac{1}{2M_1^2} y^R_{tr} y^{L*}_{sp} +\frac{1}{2M_2^2} y^{(u)*}_{tp} y^{(q)}_{sr}  \,, &
\cwc{\ell equ}{}[(3)][prst] &= -\frac{1}{ 8M_1^2} y^R_{tr} y^{L*}_{sp} +  \frac{1}{ 8M_2^2} y^{(u)*}_{tp} y^{(q)}_{sr}  \,,  \nn
\cwc{\ell q}{}[(1)][prst] &= \frac{1}{4M_1^2} y^L_{tr} y^{L*}_{sp} +  \frac{3}{4M_3^2} y^S_{tr} y^{S*}_{sp} \,, &
\cwc{\ell q}{}[(3)][prst] &=  -\frac{1}{4M_1^2} y^L_{tr} y^{L*}_{sp} +\frac{1}{4M_3^2} y^S_{tr} y^{S*}_{sp} \,, \nn
\cwc{q e}{}[][prst]  &= -\frac{1}{2M_2^2} y^{(q)}_{pt} y^{(q)*}_{rs} \,, &\cwc{l u}{}[][prst]  &= -\frac{1}{2M_2^2} y^{(u)}_{sr} y^{(u)*}_{tp} \,, \nn
\cwc{e u}{}[][prst] &=  \frac{1}{2M_1^2} y^R_{tr} y^{R*}_{sp} \,,
\end{align}
where $M_{1,2,3}$ are the masses of $S_1$, $R_2$ and $S_3$. The $S_1$ and $S_3$ contributions agree with the results in Ref.~\cite{Gherardi:2020det} and the ones for $R_2$ with Refs.~\cite{Feruglio:2018fxo,deBlas:2017xtg}. From Eq.~\eqref{eq:LQtree} it can  already be seen that $S_1$ and $R_2$ give a large contribution to the magnetic moments through the tree-level matching to $\cwc{\ell equ}{}[(3)][\ell\ell 33]$ if they couple to $t$ quarks and the corresponding leptons. Finally, the QCD corrections to Eq.~\eqref{eq:LQtree} can be shown to be about 10\% \cite{Aebischer:2018acj}.

The one-loop diagrams shown in Fig.~\ref{g2MU} generate contributions to the leptonic dipole operators
\begin{align}
\cwc{eB}{}[][pr] &= \frac{g_1 N_c}{16 \pi^2 } \biggl\{ - \frac{y^{L*}_{sp}  y^R_{tr}}{4 M_1^2}    \big[Y_u\big]_{ts} \left[\left(\mathsf{y}_q+\mathsf{y}_u\right)\log\frac{\mu^2}{M_1^2}+ \frac{3}{2} \mathsf{y}_q+\frac{1}{2}\mathsf{y}_u -\mathsf{y}_e\right] \nn
& +    \left[ \frac{y^{L*}_{sp}  y^L_{st} }{24 M_1^2} + \frac{ y^{S*}_{sp}  y^S_{st} }{8 M_3^2} \right]  \big[Y^*_e\big]_{rt} (\mathsf{y}_e +\mathsf{y}_u  + 2 \mathsf{y}_q )
+ \frac{y^{R*}_{ts} y^R_{tr}}{24  M_1^2}  \big[Y^*_e\big]_{sp}  (\mathsf{y}_e +3 \mathsf{y}_u ) \nn
&+  \frac{y^{{(u)}*}_{sp}  y^{(q)}_{tr}}{4M_2^2}    \big[Y_u\big]_{st} \left[ \left(\mathsf{y}_q+\mathsf{y}_u\right)\log\frac{\mu^2}{M_2^2} + \frac{1}{2} \mathsf{y}_q + \frac{3}{2}\mathsf{y}_u + \mathsf{y}_e\right] \nn
& +    \frac{y^{(u)*}_{sp}  y^{(u)}_{st} }{24 M_2^2}    \big[Y^*_e\big]_{rt}  (\mathsf{y}_e -\mathsf{y}_q  - 2 \mathsf{y}_u )
+ \frac{y^{(q)*}_{ts} y^{(q)}_{tr}}{24 M_2^2}  \big[Y^*_e\big]_{sp}  (2\mathsf{y}_e -6 \mathsf{y}_q )  \biggr\} \,, \nn
\cwc{eW}{}[][pr] &= \frac{g_2 N_c}{16 \pi^2 }  \biggl\{ \frac{y^{L*}_{sp} y^R_{tr} }{8 M_1^2}  \big[Y_u\big]_{ts} \left[\log\frac{\mu^2}{M_1^2}+ \frac{3}{2} \right] + \left[- \frac{y^{L*}_{sp}  y^L_{st}}{24 M_1^2} + \frac{y^{S*}_{sp}  y^S_{st}}{8 M_3^2} \right]  \big[Y^*_e\big]_{rt}   \nn
& -\frac{y^{(u)*}_{sp} y^{(q)}_{tr} }{8 M_2^2}  \big[Y_u\big]_{st} \left[\log\frac{\mu^2}{M_2^2}+ \frac12  \right] + \frac{y^{(u)*}_{sp}  y^{(u)}_{st}}{48 M_2^2}  \big[Y^*_e\big]_{rt}   \biggr\}   \,,
\end{align}
where the results for $S_1$ and $R_2$ are consistent with Ref.~\cite{Dekens:2018bci}.\footnote{We follow the notation in Refs.~\cite{Jenkins:2013zja,Jenkins:2013wua,Alonso:2013hga}, $\mathsf{y}_l=-1/2$, $\mathsf{y}_e=-1$, $\mathsf{y}_q=1/6$, $\mathsf{y}_u=2/3$, $\mathsf{y}_d=-1/3$.
Note that $\mathsf{y}_e +\mathsf{y}_u  + 2 \mathsf{y}_q =0$.} The $S_1$ and $S_3$ contributions were computed in Ref.~\cite{Gherardi:2020det}, and we have some minor differences with their result.  Although all three leptoquarks contribute to $a_{l}$ via  dipole operators, only the $S_1$  and $R_2$ leptoquarks give rise to contributions that are chirally enhanced. These leptoquark contributions are proportional to the up-type Yukawa couplings, $Y_u$, and provide a large contribution from $t$-quark loops. $S_3$ only contributes proportional to the lepton Yukawa couplings $Y_e$.

Between $\mu = M_{1,2}$ and the electroweak scale the semi-leptonic tensor operators, with coefficients $\cwc{\ell equ}{}[(3)][]$ given in Eq.~\eqref{eq:LQtree}, mix into the dipole operators via RG running. At the electroweak scale  $\cwc{eB}{}[][]$ and $\cwc{eW}{}[][]$ match onto $\lwc{e\gamma}{}[][]$ at tree level, while $\cwc{\ell equ}{}[(3)][]$ contributes at one loop. Focusing on the enhanced contributions involving the top quark,
one finds the fixed-order result~\cite{Djouadi:1989md,Cheung:2001ip,Lavoura:2003xp,Dorsner:2020aaz}
\begin{figure}[t]
\begin{center}
	\scriptsize
	\begin{fmfgraph*}(120,70)
		\fmftop{t1} \fmfbottom{b1,b2}
		\fmf{quark,label=$\mu$,label.side=right,tension=3}{b1,v1}
		\fmf{quark,label=$t$,label.side=left}{v2,v1}
		\fmf{quark,label=$\mu$,label.side=right,tension=3}{v2,b2}
		\fmf{photon,label=$\gamma$,tension=2}{t1,v3}
		\fmf{scalar,right=0.45,tension=0.75,label=$S$,label.side=right}{v2,v3,v1}
		\fmfdot{v1,v2,v3}
	\end{fmfgraph*} \hspace{1cm}
	\begin{fmfgraph*}(120,70)
		\fmftop{t1} \fmfbottom{b1,b2}
		\fmf{quark,label=$\mu$,label.side=right,tension=3}{b1,v1}
		\fmf{scalar,label=$S$,label.side=left}{v2,v1}
		\fmf{quark,label=$\mu$,label.side=right,tension=3}{v2,b2}
		\fmf{photon,label=$\gamma$,tension=2}{t1,v3}
		\fmf{quark,right=0.45,tension=0.75,label=$t$,label.side=right}{v2,v3,v1}
		\fmfdot{v1,v2,v3}
	\end{fmfgraph*} \\[0.5cm]
	\begin{fmfgraph*}(120,70)
		\fmftop{t1} \fmfbottom{b1,b2}
		\fmf{quark,label=$\mu$,label.side=right,tension=3}{b1,v1}
		\fmf{quark,label=$t$,label.side=right}{v1,v2}
		\fmf{quark,label=$\mu$,label.side=right,tension=3}{v2,b2}
		\fmf{photon,label=$\gamma$,tension=2}{t1,v3}
		\fmf{scalar,right=0.45,tension=0.75,label=$R$,label.side=right}{v2,v3,v1}
		\fmfdot{v1,v2,v3}
	\end{fmfgraph*} \hspace{1cm}
	\begin{fmfgraph*}(120,70)
		\fmftop{t1} \fmfbottom{b1,b2}
		\fmf{quark,label=$\mu$,label.side=right,tension=3}{b1,v1}
		\fmf{scalar,label=$R$,label.side=left}{v2,v1}
		\fmf{quark,label=$\mu$,label.side=right,tension=3}{v2,b2}
		\fmf{photon,label=$\gamma$,tension=2}{t1,v3}
		\fmf{quark,left=0.45,tension=0.75,label=$t$,label.side=left}{v1,v3,v2}
		\fmfdot{v1,v2,v3}
	\end{fmfgraph*}
\end{center}
\caption{The one-loop leptoquark contributions to the dipole moment. The upper graphs are from the singlet $S_1$ and triplet $S_3$ leptoquarks, and the lower graphs from the doublet $R_2$ leptoquark.
\label{g2MU}}
\end{figure}
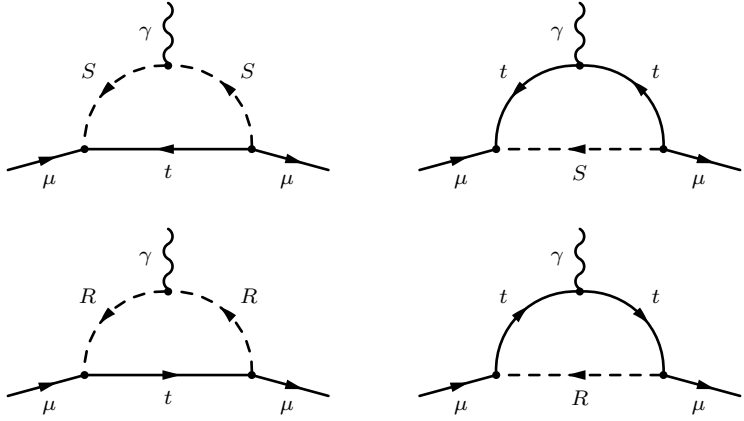
\begin{align}
\Delta a_\ell &= -  \frac{3}{8\pi^2}  \frac{m_l m_t}{M_1^2} \Re \left[ y^{L*}_{t\ell}y^R_{t\ell} \right] \left(\frac{7}{6}+\frac{2}{3}\log \frac{m_t^2}{M_1^2}\right) \nn
& +   \frac{3}{8\pi^2}  \frac{m_l m_t}{M_2^2} \Re \left[ y^{(u)*}_{t\ell}y^{(q)}_{t\ell} \right] \left(\frac{1}{6}+\frac{2}{3}\log \frac{m_t^2}{M_2^2}\right)\,,
\end{align}
which gives for the muon
\begin{align}\label{eq:LQnumeric}
\Delta a_\mu &=  \left[ 8.2 - 4.3 \log \left(\frac{(1\TeV)^2}{M_1^2} \right) \right] \times 10^{-7} \ \frac{(1\TeV)^2}{M_1^2}  \Re \left[ y^{L*}_{t\mu}y^R_{t\mu} \right] \nn
& - \left[1.5  - 0.4  \log \left(\frac{(1\TeV)^2}{M_2^2} \right)  \right] \times 10^{-6} \ \frac{(1\TeV)^2}{M_2^2}  \Re \left[ y^{(u)*}_{t\mu}y^{(q)}_{t\mu} \right] \,.
\end{align}
Including the RG resummation at leading log, in addition to the above fixed-order result, has the effect of decreasing the contributions in Eq.\ \eqref{eq:LQnumeric} by $20-30\%$ for masses $M_{1,2}$ between $1-10$ TeV.
Eq.\ \eqref{eq:LQnumeric} thus implies that the $a_\mu$ discrepancy can be explained using $M_1$ in the few to $10$ TeV range and leptoquark Yukawa couplings close to unity.


The experimental upper limit $\mathrm{Br}(\mu\to e\gamma) < 4.2 \times 10^{-13}$ \cite{TheMEG:2016wtm} severely constrains explanations of  $(g-2)_{e,\mu}$  that use leptoquarks. In particular, for any leptoquark explanation that induces both $(g-2)_e$ and $(g-2)_\mu$ through top-quark loops, the predicted $\mathrm{Br}(\mu\to e\gamma) $  is too large. Consider the $S_{1}$ scenario as an example. Assuming that the $S_1$ leptoquark only couples the electron and muon to the top quark, the chirally enhanced contributions to $\Delta a_\mu$ and   $\Delta a_e$ are
\begin{eqnarray}
\Delta a_\mu &= &-  \frac{3}{8\pi^2}  \frac{m_{\mu} m_t}{M_1^2} \Re \left[ y^{L*}_{t\mu}y^R_{t\mu} \right] \left(\frac{7}{6}+\frac{2}{3}\log \frac{m_t^2}{M_1^2}\right)\,, \nn
\Delta a_e &= &-  \frac{3}{8\pi^2}  \frac{m_e m_t}{M_1^2} \Re \left[ y^{L*}_{te}y^R_{te} \right] \left(\frac{7}{6}+\frac{2}{3}\log \frac{m_t^2}{M_1^2}\right)\,.
\end{eqnarray}
In terms of these quantities, a lower limit on the branching ratio of  $\mu\to e\gamma$ is given by \cite{Lavoura:2003xp,Crivellin:2018qmi,Dorsner:2020aaz,Bigaran:2020jil}
\begin{equation}
{\rm BR}(\mu\to e \gamma) \geq \frac{\tau_{\mu}\alpha m_{\mu}^3}{16}\left( \frac{\Delta a_{e}^{2}}{m_{e}^{2} \xi^{2}} +\frac{\Delta a_{\mu}^{2}}{m_{\mu}^{2} }\xi^{2}   \right)\,,
\end{equation}
where $\alpha$ is the fine-structure constant, $\tau_{\mu}$ the muon lifetime, and $\xi= \big|\frac{y^R_{te}}{y^R_{t\mu}}\big|$.  The minimum value of the above expression can be obtained by minimizing with respect to $\xi^{2}$. Using the observed values of $\Delta a_{e/\mu}$ we obtain
\begin{equation}
{\rm BR}(\mu\to e \gamma)_\mathrm{min} = \frac{\tau_{\mu}\alpha m_{\mu}^3}{8}\frac{|\Delta a_{e}\Delta a_{\mu}|}{m_{e}m_{\mu}}=1.5\times 10^{-4}\,,
\end{equation}
with $\Delta a_e = \Delta a_e^\mathrm{Cs}$ and $\tau_{\mu}=2.2~ \mu s=3.3\times 10^{18}~{\rm GeV}^{-1}$. Therefore, it is not possible to satisfy the constraints from $\mu\to e \gamma$ if one assumes that  $\Delta a_{e}$ and $\Delta a_{\mu}$ are both due to loops involving only the top and a single leptoquark~\cite{Crivellin:2018qmi}.

\begin{figure}[t]
\centering
	\scriptsize
	\mbox{}\\[-0.5cm]
	\begin{fmfgraph*}(80,80)
		\fmftop{t1} \fmfbottom{b1,b2} \fmfleft{l1} \fmfright{r1}
		\fmf{quark,label=$\mu$,label.side=right,tension=3}{b1,v1}
		\fmf{quark}{v1,v2}
		\fmf{quark,label=$\mu$,label.side=right,tension=3}{v2,b2}
		\fmf{photon,label=$\gamma$,tension=2}{t1,v3}
		\fmf{photon,label.side=right,label=$\gamma$}{v2,v4}
		\fmf{dashes,label.side=left,label=$H^0_i$}{v1,v5}
		\fmf{quark,left=0.6,tension=0.75,label=$f$,label.side=left}{v3,v4}
		\fmf{quark,left=0.6,tension=0.75}{v4,v5,v3}
		\fmf{phantom}{l1,v5}
		\fmf{phantom}{v4,r1}
		\fmfdot{v1,v2,v3,v4,v5}
	\end{fmfgraph*}  \hspace{1cm}
	\begin{fmfgraph*}(80,80)
		\fmfset{arrow_len}{2.5mm}
		\fmftop{t1} \fmfbottom{b1,b2} \fmfleft{l1} \fmfright{r1}
		\fmf{quark,label=$\mu$,label.side=right,tension=3}{b1,v1}
		\fmf{quark}{v1,v2}
		\fmf{quark,label=$\mu$,label.side=right,tension=3}{v2,b2}
		\fmf{photon,label=$\gamma$,tension=2}{t1,v3}
		\fmf{photon,label.side=right,label=$\gamma$}{v2,v4}
		\fmf{dashes,label.side=left,label=$H^0_i$}{v1,v5}
		\fmf{scalar,left=0.6,tension=0.75,label=$H^\pm$,label.side=left}{v3,v4}
		\fmf{scalar,left=0.6,tension=0.75}{v4,v5,v3}
		\fmf{phantom}{l1,v5}
		\fmf{phantom}{v4,r1}
		\fmfdot{v1,v2,v3,v4,v5}
	\end{fmfgraph*}  \hspace{1cm}
	\begin{fmfgraph*}(80,80)
		\fmfset{arrow_len}{2.5mm}
		\fmftop{t1} \fmfbottom{b1,b2} \fmfleft{l1} \fmfright{r1}
		\fmf{quark,label=$\mu$,label.side=right,tension=3}{b1,v1}
		\fmf{quark}{v1,v2}
		\fmf{quark,label=$\mu$,label.side=right,tension=3}{v2,b2}
		\fmf{photon,label=$\gamma$,tension=2}{t1,v3}
		\fmf{photon,label.side=right,label=$\gamma$}{v2,v4}
		\fmf{dashes,label.side=left,label=$H^0_i$}{v1,v5}
		\fmf{photon,left=0.6,tension=0.75,label=$W^\pm$,label.side=left}{v3,v4}
		\fmf{photon,left=0.6,tension=0.75}{v4,v5,v3}
		\fmf{phantom}{l1,v5}
		\fmf{phantom}{v4,r1}
		\fmfdot{v1,v2,v3,v4,v5}
	\end{fmfgraph*}
\caption{Two Loop Barr-Zee diagrams contributing to the lepton dipole moments}
\label{BZg2mu}
\end{figure}
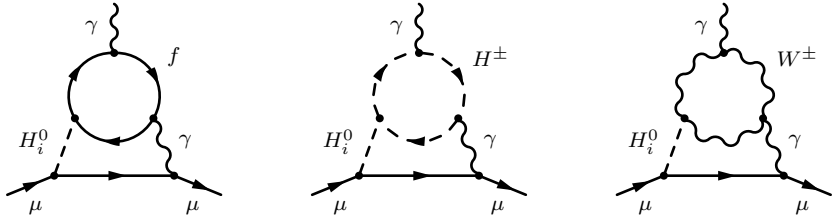
Finally, we note that two-Higgs-doublet models (2HDM) are also popular explanations of the $\Delta a_{\mu}$ anomaly, where the dominant contribution usually is from two-loop Barr--Zee diagrams shown in Fig.~\ref{BZg2mu}.
These models often have a light $CP$-odd scalar below the electroweak scale and could be described by a two-loop matching onto LEFT~\cite{Davidson:2016utf}. A detailed analysis of these models is beyond the scope of the present paper.


\section{Conclusions}
\label{sec:Conclusions}

We have analyzed the lepton magnetic and electric dipole moments in a model-independent way in terms of LEFT and SMEFT. Our main results are expressions for $\Delta a_{e,\mu}$ as a function of the LEFT and SMEFT Wilson coefficients that include one-loop running and matching effects. 
We also consider the contributions of semileptonic operators involving light quarks, which get modified by the strong interactions at low energies. We have parameterized these effects in terms of non-perturbative parameters of $\O(1)$, whose precise determination requires a non-perturbative matching calculation with lattice QCD.
Along with the current experimental measurements,  our results place strong constraints on the Wilson coefficients in these effective field theories. We find that the current muon anomaly $\Delta a_\mu \sim 2 \times 10^{-9}$ can only be generated by a very limited set of higher-dimensional operators, especially if the BSM physics scale is above a TeV. In this case, the only interactions that can produce a large enough $\Delta a_\mu$ are the dipole operators, $C_{eW}$ and $C_{eB}$, and the semileptonic four-fermion operators $\cwc{lequ}{(3)}[][2222]$ and $\cwc{lequ}{(3)}[][2233]$.
These operators need to be induced at one loop and tree level, respectively, to produce a large enough $\Delta a_\mu$. Even if there is BSM physics below the electroweak scale, only a few other semileptonic operators contribute (given in Eq.~\eqref{eq:amu60}).


	\section*{Acknowledgements}
	\addcontentsline{toc}{section}{\numberline{}Acknowledgements}

	We thank M.~Hoferichter and J.~Virto for useful discussions. Financial support by the DOE (Grant No.\ DE-SC0009919) is gratefully acknowledged.
	J.\,A.\ acknowledges financial support from the Swiss National Science Foundation (Project No.\ P400P2\_183838). D.\,S.\ is   supported by the National Science Foundation under Grant No.\ PHY-1915147.


	\appendix

	\end{myfmf}
	
	\interlinepenalty=10000

	\addcontentsline{toc}{section}{\numberline{}References}
	\bibliographystyle{JHEP}
	\bibliography{Literature}

\end{document}